# Nonlinear charge transport in highly polar semiconductors: GaN, AlN, InN and GaAs

CLÓVES G RODRIGUES [1]    and ROBERTO LUZZI[2]

[1]School of Exact Sciences and Computing, Pontifical Catholic University of Goiás, CP 86, 74605-010 Goiânia, Goiás, Brazil
[2]Condensed Matter Physics Department, Institute of Physics "Gleb Wataghin" State University of Campinas-Unicamp, 13083-859 Campinas, SP, Brazil
*Corresponding author. E-mail: cloves@pucgoias.edu.br

**Abstract.** In this paper, we present a collection of results focussing on the transport properties of doped direct-gap inverted-band highly polar III-nitride semiconductors (GaN, AlN, InN) and GaAs in the transient and steady state, calculated by using nonlinear quantum kinetic theory based on a non-equilibrium statistical ensemble formalism (NESEF). In the present paper, these results are compared with calculations using Monte Carlo modelling simulations and experimental measurements. Both *n*-type and *p*-type materials, in the presence of intermediate to high electric fields, are considered for several temperatures and carrier concentrations. The agreement between the results obtained using nonlinear quantum kinetic theory, with those of Monte Carlo calculations and experimental data is remarkably good, thus satisfactorily validating the NESEF.

## 1. Introduction

In a general article in the October 2000 issue of *Physics Today* [1] it was stated that: "The recent achievement of compact blue-emitting gallium nitride semiconductor lasers is likely to have far-reaching technological and commercial effects. The lasers' short wavelengths – around 400 nm, half that of gallium arsenide-based lasers – permit higher spatial resolution in applications such as optical storage and printing. And the high photon energy will open up new applications for these inexpensive, compact light sources. An aesthetic satisfaction with these devices stems from finally extending the existing and mature semiconductor laser technology for the near-infrared and red to encompass the "new frontier" of blue and near-ultraviolet regions, thereby bridging the entire visible spectrum".

Moreover, the same article indicated that one major commercial impact of blue diode lasers will be in high-density optical storage involving digital versatile disks as well as in colour projection displays and high-resolution laser printers. In addition, a significant technological opportunity is offered by GaN-based light-emitting diodes (LEDs), in which incoherent light is produced by spontaneous, as opposed to stimulated, emissions.

Examples of commercial uses include green LEDs for traffic lights, violet LEDs that can be combined with phosphors to produce white light, as a replacement for the ubiquitous incandescent light bulb, and blue and green nitride LEDs that can be integrated into large-scale outdoor displays.

Of further note is an article in *Scientific American* (August 2000) [2], which states that "For more than 25 years, LEDs were like a third of a rainbow. Red, orange, yellow, and the yellowish green were all you could get. Engineers wanted blue and true green because with these colors, along with the red they already had, they could built fabulous things, such as white-light-emitting devices as much as 12 times more efficient and longer-lasting than ordinary light bulb. Small wonder, then, that analysts say LEDs are poised to revolutionize the lighting industry and more beyond their familiar role as mere indicator light. In the mean time, colored LEDs are being deployed as traffic lights and in displays, the biggest being the eight-story-tall Nasdaq display in New York, in City's Time Square. And a blue semiconductor laser, similar to a LED, will soon quadruple the storage of capacity of DVD and CD players and the resolution of laser prints. Most of the milestones on the way to these optoelectronics triumphs took place, oddly enough, on the island of Shikoku, something of a backwater in the Japanese chain [from Nichia Corporation, a once obscure Japanese maker of phosphors for cathode-ray tubes and fluorescent lights]."

The successful commercialisation of III-nitrides-based light-emitting diodes in the blue–yellow range, and the development of injection lasers and ultraviolet detectors, initially drove the focus of basic research on these wide-gap semiconductors in order to understand their optical properties [3–6]. However, their favourable properties for the implementation of electronic power devices, high-performance and high-frequency transistors [7–11] as well as their superior characteristics over those based on silicon [12,13] were soon recognised. The efforts undertaken to investigate the steady-state transport properties of III-nitrides, looking for a better determination of bulk material parameters and band structure [14–19], are relevant for establishing their figure of merit with respect to electronic devices with certain confidence.

These properties have been considered since the end of the 20th century and the beginning of the 21st century [1,2,2–24,26–36]. During the first decade of the current century (and new millennium), development of some commercial applications has resulted. In early research, it was also noted that significant research opportunities are arising from a plethora of poorly understood microscopic issues in the underlying material system, which include such fundamental properties as charge control, transport and formation of optical gain for stimulated emission [1,2,2–24,26–36].

Focus on the improvement of III-nitrides-based devices through the use of submicron channels has stimulated research on their transient-transport properties, because in this case, the carriers may not attain the steady-state transport regime. The possibility of transient ballistic transport in GaN was demonstrated to occur when electric fields greater than 140 kV/cm are applied [37], and investigations of the transient transport regime have indicated the possibility of an overshoot effect in both the electron drift velocity and mean energy [38–40]. The explanation for the existence of the overshoot effect in the III-nitrides, when intervalley scattering is negligible, was considered to be due to the interplay of energy and momentum relaxation times. No overshoot occurs when the momentum relaxation time, which is less than the energy relaxation time shortly after application of the electric field, becomes predominantly larger than the other. On the other hand, the overshoot occurs at intermediate to high fields when the energy relaxation time remains larger than the momentum relaxation time [41,42].

The detailed theoretical derivation and analysis of nonlinear transport in semiconductors requires the proper specification of the appropriate non-equilibrium thermodynamic state spaces, and derivation of the evolution equations for the non-equilibrium thermodynamic variables that span such space states. We shall describe how they are built and how they are applied within the non-equilibrium statistical ensemble formalism (NESEF [43]), as described in Appendix, which provides an elegant, practical and physically clear picture to evaluate irreversible processes [44]. Note that the theory of transport has been approached differently during the past century, undergoing a continuous process of improvement (the origin dates back to the fundamental work by the great Ludwig Boltzmann). Such approaches have been classified by Zwanzig [45]. At present, we mention the two front line approaches, namely, computational methods derived from non-equilibrium molecular dynamics (NMD) and nonlinear quantum kinetic theory based on NESEF described in Appendix. In particular, both approaches have advantages and disadvantages in their general formulation, but in the case of charge transport in semiconductors (the topic of our interest), they are as equivalent in the sense of providing numerical results that are very similar in several cases. Some examples are presented in [46]. The NESEF-based transport theory is based on averaging the non-equilibrium statistical ensemble of the mechanical Heisenberg equations of motion for the observables relevant to the problem at hand. Zwanzig stated that this formalism has by far the most appealing structure and may be the most effective method for dealing with nonlinear transport processes [45]. NESEF provides an elegant, practical and physically clear picture for describing irreversible processes [48], as in the case when semiconductors are far from the equilibrium state [41,42,46,49–89].

Addressing this issue, we present a collection of results dealing with transport (charge transport under intermediate to high electric fields) of gallium arsenide (GaAs) and III-nitrides: gallium nitride (GaN), aluminum nitride (AlN) and indium nitride (InN).

## 2. Theoretical background

Consider an n-doped direct-gap polar semiconductor in a state such that the extra electrons act as mobile carriers in the conduction band. We use the effective mass approximation and therefore the parabolic bands;

this implies that in explicit applications, an upper limiting value for the electric field exists that needs to be controlled so that values lower than this value allow intervalley scattering to be neglected. The Hamiltonian of the system is

$$\hat{H} = \hat{H}_0 + \hat{H}_1 + \hat{H}_{imp} + \hat{H}_{AN} + \hat{H}_{CF} + \hat{H}_{SR} , \qquad (1)$$

where

$$\hat{H}_0 = \sum_{\mathbf{k}} \frac{\hbar^2 k^2}{2m_e^*} c_{\mathbf{k}}^\dagger c_{\mathbf{k}} + \sum_{\mathbf{q}, \eta} \hbar \omega_{\mathbf{q}, \eta} \left( b_{\mathbf{q}, \eta}^\dagger b_{\mathbf{q}, \eta} + 1/2 \right) \qquad (2)$$

is the Hamiltonian of the free electrons and phonons, and

$$\hat{H}_1 = \sum_{\mathbf{k}, \mathbf{q}} \sum_{\eta, \ell} \left( M_\eta^\ell(\mathbf{q}) b_{\mathbf{q}, \eta} c_{\mathbf{k}+\mathbf{q}, a}^\dagger c_{\mathbf{k}} \right.$$
$$\left. + M_\eta^{\ell *}(\mathbf{q}) b_{\mathbf{q}, \eta}^\dagger c_{\mathbf{k}+\mathbf{q}} c_{\mathbf{k}}^\dagger \right) , \qquad (3)$$

contains the electron–phonon interactions. In the equations above, $c(c^\dagger)$ and $b(b^\dagger)$ are the annihilation (creation) operators in electron states $|\mathbf{k}\rangle$ and of phonons in mode $|\mathbf{q}\rangle$ associated with $\eta =$ LO, AC (for longitudinal-optical and acoustical phonons, respectively). $M_\eta^\ell(\mathbf{q})$ are the matrix elements indicating the type of interaction between carriers and $\eta$-type phonons, with supraindex $\ell$ indicating the type of interaction (polar, deformation potential, piezoelectric, etc.). The third contribution, $\hat{H}_{imp}$, refers to the interaction of carriers with impurities (see, for example, [90,91]). Moreover, $\hat{H}_{AN}$ represents the anharmonic interactions in the phonon system, and

$$\hat{H}_{CF} = \sum_i e \mathcal{E} \cdot \mathbf{r}_i \qquad (4)$$

defines the interactions of the electrons (with charge $-e$ and position $\mathbf{r}_i$) with an electric field of intensity $\mathcal{E}$. The interaction of the system with an external reservoir is included in $\hat{H}_{SR}$ in eq. (1); the reservoir is considered as an ideal one, which is satisfactory in most cases, and therefore its macroscopic (thermodynamic) state is characterised by a canonical statistical distribution with temperature $T_0$.

Now, consider the non-equilibrium thermodynamic state of the plasma: the presence of the electric field changes the energy of the electrons (they acquire energy in excess of equilibrium), and these carriers continue transferring this excess energy to the lattice such that an electrical current (flux of electrons) follows. Thus, we need to choose the basic variables

$$\{ E_e(t), N_e(t), \mathbf{P}_e(t), E_{LO}(t), E_{AC}(t), E_R \} , \qquad (5)$$

which are, respectively, the energy, number and linear momentum of the electrons, the energies of the LO and AC phonons (noting that electrons in conduction states with $s$-type symmetry do not interact with transverse-optical phonons) and the energy of the reservoir. The corresponding dynamical quantities are defined as

$$\{ \hat{H}_e, \hat{N}_e, \hat{\mathbf{P}}_e, \hat{H}_{LO}, \hat{H}_{AC}, \hat{H}_R \} , \qquad (6)$$

which are the Hermitian operators for the partial Hamiltonians, the electron number and the linear momentum.

According to the NESEF and the accompanying irreversible thermodynamics principles [44,47,92–94], the non-equilibrium thermodynamic state of the system can completely, and alternatively to the description provided by the variables of eq. (5), be characterised by a set of intensive non-equilibrium thermodynamic variables (the Lagrange multipliers provided by the variational construction of the formalism), namely

$$\{ F_e(t), F_{n_e}(t), \mathbf{F}_e(t), F_{LO}(t), F_{AC}(t), \beta_0 \} . \qquad (7)$$

The variables in eq. (7) are present in the auxiliary statistical operator introduced by the formalism, which in this case is given by

$$\bar{\rho}(t, 0) = \rho_R \times \exp \left\{ -\phi(t) - F_e(t)\hat{H}_e \right.$$
$$-F_{n_e}(t)\hat{N}_e - \mathbf{F}_e(t) \cdot \hat{\mathbf{P}}_e - F_{LO}(t)\hat{H}_{LO}$$
$$\left. -F_{AC}(t)\hat{H}_{AC} \right\} , \qquad (8)$$

where $\rho_R$ is the canonical distribution of the reservoir at temperature $T_0$. Note that the operator in eq. (8) is not the statistical operator describing the macroscopic state of the system, which is a superoperator of this system [43] and $\phi(t)$ (playing the role of the logarithm of a non-equilibrium partition function) ensures the normalisation of $\bar{\rho}(t, 0)$.

The non-equilibrium statistical operator, in terms of the auxiliary one in eq. (8), is given by

$$\rho_\varepsilon(t) = \exp \left\{ -\ln \bar{\rho}(t, 0) \right.$$
$$\left. + \int_{-\infty}^t dt' e^{\varepsilon(t'-t)} \frac{d}{dt'} \ln \bar{\rho}(t', t' - t) \right\} ,$$

where

$$\bar{\rho}(t', t' - t) = e^{-(t'-t)\hat{H}/i\hbar} \bar{\rho}(t', 0) e^{(t'-t)\hat{H}/i\hbar}$$

and $e^{\varepsilon(t'-t)}$ is a kernel (Abel's kernel in the theory of convergence of integral transforms) that ensures irreversible evolution from the initial condition of preparation of an irreversibility introduced in the interventionist approach in logic taken according to the Krylov–Bogoliubov 'jolting' proposition [43]. Quantity $\varepsilon$ is a positive infinitesimal that goes to zero after the calculation of averages.

The intensive non-equilibrium thermodynamic variables of eq. (7) are usually redefined as

$$F_e(t) = \beta_e^*(t) = \frac{1}{k_B T_e^*(t)} , \qquad (9a)$$

$$F_{n_e}(t) = -\beta_e^*(t)\mu_e^*(t), \tag{9b}$$

$$\mathbf{F}_e(t) = -\beta_e^*(t)\mathbf{v}_e(t), \tag{10}$$

$$F_{LO}(t) = \beta_{LO}^*(t) = \frac{1}{k_B T_{LO}^*(t)}, \tag{11a}$$

$$F_{AC}(t) = \beta_{AC}^*(t) = \frac{1}{k_B T_{AC}^*(t)}, \tag{11b}$$

where $k_B$ is the usual Boltzmann constant and $\beta_0$ in eq. (7) is defined as

$$\beta_0 = \frac{1}{k_B T_0}.$$

Equations (9)–(11) introduce the so-called non-equilibrium temperatures (or quasitemperatures), $T_e^*(t)$, $T_{LO}^*(t)$, $T_{AC}^*(t)$, of electrons and phonons, the quasi-chemical potential, $\mu_e^*$, and the drift velocity, $\mathbf{v}_e(t)$ [92].

We now derive the equations of evolution for the basic variables of eq. (5). If we generically refer to the variables of eq. (5) as $Q_j(t)$, and as $\hat{P}_j$ to those of eq. (6), the general form of the equations of evolution is [95–97]

$$\frac{d}{dt}Q_j(\mathbf{r}, t) \cong J_j^{(0)}(\mathbf{r}, t) + J_j^{(1)}(\mathbf{r}, t) + J_j^{(2)}(\mathbf{r}, t), \tag{12}$$

where

$$J_j^{(0)}(\mathbf{r}, t) = \frac{1}{i\hbar}\mathrm{Tr}\big\{[\hat{P}_j(\mathbf{r}), \hat{H}_0]\bar{\rho}(t, 0)\big\}, \tag{13}$$

$$J_j^{(1)}(\mathbf{r}, t) = \frac{1}{i\hbar}\mathrm{Tr}\big\{[\hat{P}_j(\mathbf{r}), \hat{H}_1]\bar{\rho}(t, 0)\big\} \tag{14}$$

and

$$J_j^{(2)}(\mathbf{r}, t) = -\frac{1}{\hbar^2}\lim_{\epsilon \to +0}\int_{-\infty}^0 dt' e^{\epsilon t'}$$
$$\times \mathrm{Tr}\bigg\{\Big(\big[\hat{H}_1(t')_0, [\hat{H}_1, \hat{P}_j(\mathbf{r})]\big]$$
$$+ i\hbar\sum_k[\hat{H}_1(t')_0, \hat{P}_k(\mathbf{r})]\frac{\partial J_k^{(1)}(\mathbf{r}, t)}{\partial Q_k(\mathbf{r}, t)}\Big)\bar{\rho}(t, 0)\bigg\}, \tag{15}$$

where the nought subindex denotes evolution in the interaction representation, that is, the evolution of $\hat{H}_1$ under $\hat{H}_0$ alone.

Proceeding with the calculations we obtain the corresponding set of equations of evolution, namely

$$\frac{d}{dt}E_e(t) = -\frac{e}{m_e^*}\boldsymbol{\mathcal{E}} \cdot \mathbf{P}_e(t) + J_{E_{LO}}^{(2)}(t) + J_{E_{AC}}^{(2)}(t), \tag{16}$$

$$\frac{d}{dt}N_e(t) = 0, \tag{17}$$

$$\frac{d}{dt}\mathbf{P}_e(t) = -nVe\boldsymbol{\mathcal{E}} + \mathbf{J}_{\mathbf{P}_e}^{(2)}(t) + \mathbf{J}_{\mathbf{P}_E,\mathrm{imp}}^{(2)}, \tag{18}$$

$$\frac{d}{dt}E_{LO}(t) = -J_{E_{LO}}^{(2)}(t) + J_{LO,\mathrm{an}}^{(2)}(t), \tag{19}$$

$$\frac{d}{dt}E_{AC}(t) = -J_{E_{AC}}^{(2)}(t) - J_{LO,\mathrm{an}}^{(2)}(t) + J_{AC,\mathrm{dif}}^{(2)}(t). \tag{20}$$

We analyse these equations term by term. In eq. (16) the first term on the right accounts for the rate of energy transferred from the electric field to the carriers. The second and third terms account for the transfer of the excess energy of the carriers to the LO and AC phonons, respectively, and are given by

$$J_{E_\eta}^{(2)} = \frac{2\pi}{\hbar}\sum_{\mathbf{k},\mathbf{q},\ell}|M_\eta^\ell(\mathbf{q})|^2(\epsilon_{\mathbf{k}+\mathbf{q}}^e - \epsilon_{\mathbf{k}}^e)\big[\nu_{\mathbf{q},\eta}(t)f_{\mathbf{k}}^e(t)$$
$$\times \big(1 - f_{\mathbf{k}+\mathbf{q}}^e(t)\big) - f_{\mathbf{k}+\mathbf{q}}^e(t)\big(1 + \nu_{\mathbf{q},\eta}(t)\big)$$
$$\times \big(1 - f_{\mathbf{k}}^e(t)\big)\big]\delta(\epsilon_{\mathbf{k}+\mathbf{q}}^e - \epsilon_{\mathbf{k}}^e - \hbar\omega_{\mathbf{q},\eta}), \tag{21}$$

where $\eta = $ LO or AC, and

$$\epsilon_{\mathbf{k}}^e = \frac{\hbar^2 k^2}{2m_e^*}, \tag{22}$$

(recalling that we are using the random phase approximation (RPA) and the effective mass approximation), and

$$f_{\mathbf{k}}^e(t) = \mathrm{Tr}\big\{c_{\mathbf{k}}^\dagger c_{\mathbf{k}}\bar{\rho}(t)\big\}$$
$$f_{\mathbf{k}}^e(t) = \frac{1}{e^{\beta_e^*(t)[(\hbar^2(\mathbf{k}-m_e^*\mathbf{v}_e(t)/\hbar)^2/2m_e^*) - \mu_e^*(t)]} + 1}, \tag{23}$$

which, calculated in terms of the variables in eq. (6), takes the form of an instantaneous Fermi–Dirac-like distribution (in time) with non-equilibrium temperature $T^*(t)$ and quasichemical potential $\mu_e^*(t)$, and drift because of the action of the electric field that can be described by the drift velocity $\mathbf{v}_e(t)$. Note that this is the result of the very rapid internal thermalisation of the electrons as a consequence of the Coulomb interactions among them. Moreover, in many cases, a quasiclassic approach is possible, that is, the condition $n\Lambda_e^3(t) \ll 1$ holds, where $\Lambda_e(t) = \hbar/\sqrt{m_e^* k_B T_e^*(t)}$ is an instantaneous non-equilibrium-thermal de Broglie wavelength. Then the distribution of eq. (23) can be approximated as follows:

$$f_{\mathbf{k}}^e(t) = 4n\Big(\frac{\pi\hbar^2}{2m_e^* k_B T_e^*(t)}\Big)^{3/2}$$
$$\times \exp\Big\{-\frac{\hbar^2\big(\mathbf{k}-m_e^*\mathbf{v}_e(t)/\hbar\big)^2}{2m_e^* k_B T_e^*(t)}\Big\}, \tag{24}$$

which resembles a time-dependent (on the evolution of the non-equilibrium state of the system) and drifted Maxwell–Boltzmann-type of distribution. Moreover,

$$\nu_{\mathbf{q},\eta} = \mathrm{Tr}\{b_{\mathbf{q},\eta}^\dagger b\bar{\rho}(t, 0)\} = \frac{1}{e^{\hbar\omega_{\mathbf{q},\eta}/k_B T_\eta^*(t)} - 1}. \tag{25}$$

Equation (17) accounts for the fact that the concentration $n$ of electrons is fixed by doping. In eq. (18) the first term on the right is the driving force generated by the presence of electric field. The second term is the rate of momentum transfer due to interactions with the phonons, and is given by

$$
\begin{aligned}
\mathbf{J}_{\mathbf{P}_e}^{(2)} &= \frac{2\pi}{\hbar} \sum_{\mathbf{k},\mathbf{q},\ell,\eta} \hbar \mathbf{q} |M_\eta^\ell(\mathbf{q})|^2 \big[ v_{\mathbf{q},\eta}(t) f_{\mathbf{k}}^e(t) \big(1 - f_{\mathbf{k}+\mathbf{q}}^e(t)\big) \\
&\quad - f_{\mathbf{k}+\mathbf{q}}^e(t) \big(1 + v_{\mathbf{q},\eta}(t)\big) \big(1 - f_{\mathbf{k}}^e(t)\big) \big] \\
&\quad \times \delta(\epsilon_{\mathbf{k}+\mathbf{q}}^e - \epsilon_{\mathbf{k}}^e - \hbar\omega_{\mathbf{q},\eta}) \,,
\end{aligned}
\tag{26}
$$

where $\eta = $ LO or AC.

In eqs (19) and (20), the first term on the right describes the rate of change of the energy of phonons due to their interactions with the electrons. More precisely, they account for the gain of the energy transferred to them from the hot carriers such that contributions $J_{E_{\mathrm{LO}}}^{(2)}$ and $J_{E_{\mathrm{AC}}}^{(2)}$ are the same as those calculated in eq. (21) for $\eta = $ LO or AC, with a change of sign. The second term in eq. (19) accounts for the rate of energy transfer from the optical phonons to the acoustic phonons, which is

$$
J_{\mathrm{LO,an}}^{(2)}(t) = - \sum_{\mathbf{q}} \hbar\omega_{\mathbf{q},\mathrm{LO}} \frac{v_{\mathbf{q},\mathrm{LO}}(t) - v_{\mathbf{q},\mathrm{LO}}^{\mathrm{AC}}}{\tau_{\mathrm{LO,an}}} \,,
\tag{27}
$$

where

$$
v_{\mathbf{q},\mathrm{LO}}^{\mathrm{AC}}(t) = \frac{1}{e^{\hbar\omega_{\mathbf{q},\mathrm{LO}}/k_B T_{\mathrm{AC}}^*(t)} - 1} \,.
\tag{28}
$$

In this equation, $\tau_{\mathrm{LO,an}}$ is the relaxation time which is obtained from the bandwidth in Raman scattering experiments, as described in [98]. Note that this contribution can be calculated using eq. (15), but there is no proper knowledge regarding the matrix element of the anharmonic interaction, which is usually left as a free parameter to be determined from the measurement of the relaxation time. The contribution $J_{\mathrm{LO,an}}^{(2)}(t)$ is the same but with opposite sign when inserted into eqs (19) and (20), indicating that energy is transferred from the optical phonons to the acoustical phonons. Finally, for the diffusion of heat from the AC phonons to the reservoir (the last term in eq. (20)),

$$
J_{\mathrm{AC,dif}}^{(2)}(t) = - \sum_{\mathbf{q}} \hbar\omega_{\mathbf{q},\mathrm{AC}} \frac{v_{\mathbf{q},\mathrm{AC}}(t) - v_{\mathbf{q},\mathrm{AC}}^{\mathrm{eq}}}{\tau_{\mathrm{AC,dif}}} \,,
\tag{29}
$$

where $\tau_{\mathrm{AC,dif}}$ is the characteristic time for heat diffusion, which depends on the particulars of the contact between the sample and reservoir [99].

We next present the detailed calculation of different scattering operators associated with the interactions between carriers and phonons. In the case of polar interactions with LO phonons, the matrix element is

$$
|M_{\mathrm{LO}}^{\mathrm{PO}}(\mathbf{q})|^2 = 2\pi\hbar^2 \frac{eE_{0e}}{V m_e^* q^2} \,,
\tag{30}
$$

where

$$
eE_{0e} = \frac{e^2 m_e^* \hbar\omega_{\mathrm{LO}}}{\hbar^2} \left( \frac{1}{\epsilon_\infty} - \frac{1}{\epsilon_0} \right)
$$

and $\varepsilon_0$ and $\varepsilon_\infty$ are the static and optical dielectric constants, respectively, $q$ is the modulus of the wave vector of the LO phonon in mode $|\mathbf{q}\rangle$, $\omega_{\mathrm{LO}}$ is the frequency of the LO phonons (dispersionless in an Einstein model) and $V$ is the volume. Using this matrix element, we find that

$$
\begin{aligned}
J_{E_e,\mathrm{PO}}^{(2)}(t) &= nV(e\omega_{\mathrm{LO}})^2 \sqrt{\frac{2m_e^*}{\pi k_B T_e^*(t)}} \left( \frac{1}{\epsilon_\infty} - \frac{1}{\epsilon_0} \right) \\
&\quad \times e^{z(t)/2} \big[ v_{\mathrm{LO}}(t) - \big(1 + v_{\mathrm{LO}}(t)\big) e^{-z(t)} \big] \\
&\quad \times K_0(z(t)/2),
\end{aligned}
\tag{31}
$$

$$
\begin{aligned}
\mathbf{J}_{\mathbf{P}_e,\mathrm{PO}}^{(2)}(t) &= \sqrt{2/9\pi}\, nV(e\omega_{\mathrm{LO}})^2 (m_e^*/k_B T_e^*(t))^{3/2} \\
&\quad \times \left( \frac{1}{\epsilon_\infty} - \frac{1}{\epsilon_0} \right) e^{z(t)/2} \big\{ \big[ v_{\mathrm{LO}}(t) \\
&\quad - \big(1 + v_{\mathrm{LO}}(t)\big) e^{-z(t)} \big] K_0(z(t)/2) \\
&\quad - \big[ v_{\mathrm{LO}}(t) + \big(1 + v_{\mathrm{LO}}(t)\big) e^{-z(t)} \big] \\
&\quad \times K_1(z(t)/2) \big\} \mathbf{v}_e(t) \,,
\end{aligned}
\tag{32}
$$

where

$$
z(t) \equiv \frac{\hbar\omega_{\mathrm{LO}}}{k_B T_e^*(t)}
$$

and

$$
v_{\mathrm{LO}}(t) = \frac{1}{e^{\hbar\omega_{\mathrm{LO}}/k_B T_{\mathrm{LO}}^*(t)} - 1} \,.
\tag{33}
$$

Recalling that the deformation potential of the optical phonon-carrier interaction has a matrix element that is null owing to symmetry conditions, this interaction is not present in the case of electrons in the conduction band [99].

In the case of the acoustic phonons [100]

$$
|M_{\mathrm{AC}}^{\mathrm{DP}}(\mathbf{q})| = \frac{\hbar E_1^2}{2\varrho V v_s} q,
\tag{34}
$$

where $\varrho$ is the density, $v_s$ is the sound velocity of propagation, which we consider to be the same for longitudinal acoustical (LA) and transverse acoustical (TA) phonons and $E_1$ is the acoustic deformation potential strength [100].

Using the matrix elements given above, we find that

$$
J_{E_e,\mathrm{DA}}^{(2)}(t) = -nV \frac{8\sqrt{2}(m_e^*)^{5/2} E_1^2}{\varrho\hbar^4 (\pi/k_B T_e^*(t))^{3/2}} \left( 1 - \frac{T_{\mathrm{AC}}^*(t)}{T_e^*(t)} \right),
\tag{35}
$$

$$\mathbf{J}^{(2)}_{P_e,\mathrm{DA}}(t) = -n\,V\,\frac{(m_e^*)^{5/2}E_1^2 k_\mathrm{B}T_{\mathrm{AC}}^*(t)\sqrt{k_\mathrm{B}T_e^*(t)}}{3(\pi/2)^{3/2}\varrho\, v_s^2\hbar^4}\mathbf{v}_e(t),$$

$$(36)$$

where we have used the following approximation:

$$v_\mathbf{q}^{\mathrm{AC}}(t) = \frac{1}{e^{\hbar v_s q/k_\mathrm{B}T_{\mathrm{AC}}^*(t)}-1} \simeq \frac{k_\mathrm{B}T_{ac}^*(t)}{\hbar v_s q},\qquad(37)$$

using a Debye model for the AC phonons. Then $\omega_{\mathbf{q},\mathrm{AC}} = v_s q$.

In the case of piezoelectric interaction, the matrix element is [100]

$$|M_{\mathrm{AC}}^{\mathrm{PZ}}(\mathbf{q})|^2 = \frac{e^2\hbar\omega_{\mathbf{q},\mathrm{AC}}\mathcal{K}^2}{2V\varepsilon_0 q^2},\qquad(38)$$

where $\mathcal{K}$ is the electromechanical coefficient given by

$$\mathcal{K}^2 = \frac{h_{\mathrm{PZ}}^2}{35\varepsilon_0}\left(\frac{12}{C_l}+\frac{16}{C_t}\right),$$

where $h_{\mathrm{PZ}}$ is the piezoelectric constant and $C_l$ and $C_t$ are the longitudinal and transverse elastic constants, respectively [101]. Using this matrix element, we find that

$$J^{(2)\mathrm{PZ}}_{E_e,\mathrm{AC}}(t) = -n\,V\,\frac{3\sqrt{2k_\mathrm{B}T_e^*(t)}(ev_s\mathcal{K})}{\hbar^2(\pi/m_e^*)^{3/2}\varepsilon_0}\left(1-\frac{T_{\mathrm{AC}}^*(t)}{T_e^*(t)}\right),$$

$$(39)$$

$$\mathbf{J}^{(2)\mathrm{PZ}}_{\mathbf{P}_e,\mathrm{AC}}(t) = -n\,V\,\frac{\sqrt{2/k_\mathrm{B}T_e^*(t)}k_\mathrm{B}T_{\mathrm{AC}}^*(t)(e\mathcal{K})^2}{\hbar^2(\pi/m_e^*)^{3/2}\varepsilon_0}\mathbf{v}_e(t).$$

$$(40)$$

Note that the second term in eq. (18) contains scattering due to impurities. In the case of carriers, this scattering provides little contribution to the change in energy, because a nearly elastic process exists (the impurity is much heavier than the light electron). In the case of carrier momentum, using the expression given by Ridley adapted from the results reported by Brooks-Herring [90,91],

$$\mathbf{J}^{(2)}_{\mathbf{P}_e,\mathrm{imp}} = -\frac{\mathbf{P}_e(t)}{\tau_{\mathrm{imp}}(t)}\qquad(41)$$

with

$$\tau_{\mathrm{imp}}(t) \simeq \frac{128\sqrt{2\pi m_e^*}(k_\mathrm{B}T_e^*(t))^{3/2}}{\mathcal{N}_I(\mathcal{Z}e^2/\epsilon_0)^2 G(t)},$$

where $\mathcal{N}_I$ is the density of impurities, $\mathcal{Z}$ indicates the units of charge of the impurities and

$$G(t) = \ln\{1+b(t)\} - \frac{b(t)}{1+b(t)},\qquad(42)$$

where

$$b(t) = \frac{24\epsilon_0 m_e^*(k_\mathrm{B}T_e^*(t))^2}{\mathcal{N}_I e^2\hbar^2}.\qquad(43)$$

Finally, note that all the collision operators $J^{(2)}$ are dependent on the intensive non-equilibrium thermodynamic variables of eqs (9)–(11), while the right-hand sides of eqs (16)–(20) depend on the extensive thermodynamic variables of eq. (5). To close the system of equations, we need to express the latter in terms of the former, that is, to couple the equations of evolution with the non-equilibrium thermodynamic equations of state [43,44,47], which in this case are given as [51]

$$E_e(t) = \sum_\mathbf{k}\frac{\hbar^2 k^2}{2m_e^*}f_\mathbf{k}^e(t) = \frac{N_e}{2}[3k_\mathrm{B}T_e^*(t)+m_e^*v_e^2(t)],$$

$$(44)$$

$$\mathbf{P}_e(t) = \sum_\mathbf{k}\hbar k f_\mathbf{k}^e(t) = N_e m_e^*\mathbf{v}_e(t),\qquad(45)$$

$$E_{\mathrm{LO}}(t) = \sum_\mathbf{q}\hbar\omega_{\mathbf{q},\mathrm{LO}}v_{\mathbf{q},\mathrm{LO}}(t) = \frac{V}{\mathcal{V}}\hbar\omega_{\mathrm{LO}}v_{\mathrm{LO}}(t),\quad(46)$$

$$E_{\mathrm{AC}}(t) = \sum_\mathbf{q}\hbar v_s q\, v_{\mathbf{q},\mathrm{AC}}(t) = \frac{V}{\mathcal{V}}3k_\mathrm{B}T_{\mathrm{AC}}^*(t),\quad(47)$$

where $\mathcal{V}$ is the volume of the unit cell, and eqs (24), (25) and (37) were used in the derivations.

We numerically solve the resulting equations of evolution for the intensive non-equilibrium thermodynamic variables, that is, eqs (16)–(20) together with eqs (44)–(47), which constitute a set of coupled highly nonlinear integrodifferential equations, in the next section for GaN, InN, AlN and GaAs. Table 1 presents the parameters for the three nitrides, GaN, AlN and InN, in comparison with GaAs. In table 1, $m_e^*$ is the electron effective mass, $m_h^*$ is the hole effective mass, $E_g$ is the band-gap energy, $\hbar\omega_{\mathrm{LO}}$ is the LO phonon energy, $\hbar\omega_{\mathrm{TO}}$ is the the transverse optical (TO) phonon energy, $\varepsilon_0$ is the static dielectric constant, $\varepsilon_\infty$ is the optical dielectric constant, $\alpha_e$ is the Fröhlich coupling for electrons, $a$ and $c$ are lattice parameters, $\varrho$ is the mass density, $C_l$ is the longitudinal elastic constant, $C_t$ is the transversal elastic constant, $v_s$ is the velocity of sound, $E_1$ is the acoustic deformation potential, $\mathcal{V}$ is the unit cell volume and $h_{\mathrm{pz}}$ is the piezoelectric constant.

## 3. Nonlinear transport

### 3.1 Transient state

In this section, we present a pair of examples for III nitrides (GaN, AlN and InN) regarding the evolution in time of the carrier's non-equilibrium temperature and drift velocity for different values of electric field intensity, and we explore the 'overshoot' phenomenon. The transient response of the system was generated by an abrupt turn-on (occurring at $t = 0$) of a uniform (and constant) electric field.

**Table 1.** Parameters for GaN, AlN, InN and GaAs.

| Parameter | GaN | AlN | InN | GaAs |
|---|---|---|---|---|
| $m_e^*$ ($m_0$) | 0.22 [102] | 0.29 [102] | 0.045 [102] | 0.067 [113] |
| $m_h^*$ ($m_0$) | 2.0 [103] | 3.53 [103] | 0.5 [111] | 0.5 [114] |
| $E_g$ (eV) | 3.5 [102] | 6.2 [102] | 0.63 [29,30,34] | 1.42 [115] |
| $\hbar\omega_{LO}$ (meV) | 92.0 [104] | 92.2 [22] | 89.0 [112] | 37.0 [113] |
| $\hbar\omega_{TO}$ (meV) | 69.4 [105] | – | – | 34.0 [113] |
| $\varepsilon_0$ | 9.50 [106] | 8.50 [22] | 15.3 [112] | 12.91 [116] |
| $\varepsilon_\infty$ | 5.35 [106] | 4.77 [22] | 8.40 [112] | 10.91 [116] |
| $\alpha_e$ | 0.43 | 0.60 | 0.13 | 0.1 |
| $a$ (Å) | 3.82 [107] | 3.11 [22] | 3.54 [112] | – |
| $c$ (Å) | 5.18 [106] | 4.98 [22] | 5.70 [112] | – |
| $\varrho$ (g/cm$^{-3}$) | 6.09 [108] | 3.23 [22] | 6.81 [112] | 5.31 [115] |
| $C_l$ ($10^{12}$ dyn/cm$^2$) | 2.66 [109] | 2.65 [22] | 2.65 [111] | 1.40 [117] |
| $C_t$ ($10^{11}$ dyn/cm$^2$) | 4.41 [109] | 4.42 [22] | 4.43 [111] | 4.87 [117] |
| $v_s$ ($10^5$ cm/s) | 4.40 | 6.04 | 4.15 | 3.86 |
| $E_1$ (eV) | 8.3 [17] | 9.5 [22] | 7.1 [112] | 7.0 [118] |
| $\mathcal{V}$ ($10^{-22}$ cm$^3$) | 3.07 [106,107] | 2.77 [22] | 4.13 [112] | 1.8 [116] |
| $h_{pz}$ (C/m$^2$) | 0.375 [17] | 0.92 [110] | 0.375 [112] | 0.16 [119] |

Figure 1 shows the evolution of the carriers' non-equilibrium temperature for several values of electric field for GaN, AlN and InN. After a transient time of the order of 200 fs, a steady state is attained. The time evolution of the drift velocity is shown in figure 2, where as shown in figure 1, a steady state follows after approximately 200 to 300 fs. The time for the electrons to attain the steady state is very similar to the time obtained previously by other researchers using different descriptions of the transient transport phenomena [38–40].

For a better analysis of the onset of electron velocity (temperature) overshoot, the evolution of the electron drift velocity and non-equilibrium temperature towards steady state is depicted in figures 3 and 4, respectively. These figures permit characterisation of the presence of an overshoot in both the electron drift velocity and non-equilibrium temperature: onset of the overshoot effect occurs at 20 kV/cm in GaN, 60 kV/cm in AlN and 5 kV/cm in InN. On the other hand, the smooth overshoot of the carriers' non-equilibrium temperature in GaN, AlN and InN, which was not reported previously, is due to the changes in the LO phonon temperature related to the carriers' excess energy dissipation, which was not considered by Caetano *et al* [39] and Foutz *et al* [40]. The electric field value we found for the onset of velocity overshoot in wurtzite GaN agrees well with that calculated by Caetano *et al* [39], who used

energy–momentum balance equations. During the transient regime, however, the mean electron energy (which is directly related to the electron non-equilibrium temperature) does not exhibit an overshoot effect in the results by Caetano *et al* [39]. On the other hand, the values of electric field for the onset of velocity overshoot in wurtzite AlN, GaN and InN are much lower than those obtained from the Monte Carlo simulations performed by Foutz *et al* [40]. This cannot be mostly attributed to the intervalley scattering they considered (which is not effective when the electric field intensity is less than 120 kV/cm for GaN and AlN, and 20 kV/cm for InN) or to the Fermi–Dirac-like distribution function we have used (after 100 fs the distribution function becomes Fermi–Dirac-like), but is principally related to the high values of electron effective masses in the *L–M* and *K* valleys that the authors assumed. In contrast to the less than 25% electron drift velocity overshoot, we have demonstrated for electric fields lower than 120 kV/cm that the intervalley scattering-related overshoot effect, as obtained by Foutz *et al* [40], is greater than 100%. Consequently, in the case of electric fields less than 120 kV/cm (GaN and AlN) or 20 kV/cm (InN), the possible contribution of the overshoot effect does not improve the performance of nitride-based heterojunction field-effect transistors (HFETs) [40].

An analysis of different pumping and relaxation channels [42] allows us to conclude that the overshoot that occurs at sufficiently high fields is a consequence of the interplay between energy and momentum relaxation times. No overshoot occurs when the momentum relaxation time, which is lower than the energy relaxation time shortly after the application of electric field, becomes predominantly larger than the other. On the other hand, the overshoot follows at intermediate to high fields when the relaxation time for energy is constantly larger than that for momentum. Moreover, it is verified that the peak in velocity follows in the time interval where the drift kinetic energy $m_e^* v^2(t)/2$ increases more rapidly than the thermal energy $k_B T_C^*(t)$, and that the peak in the non-equilibrium temperature occurs at the minimum value of the quotient between these two energies [42]. On the other hand, the smooth overshoot on the carriers' non-equilibrium temperature in GaN, AlN and InN is due to the changes in the LO-phonon temperature resulting from the carriers' excess-energy dissipation.

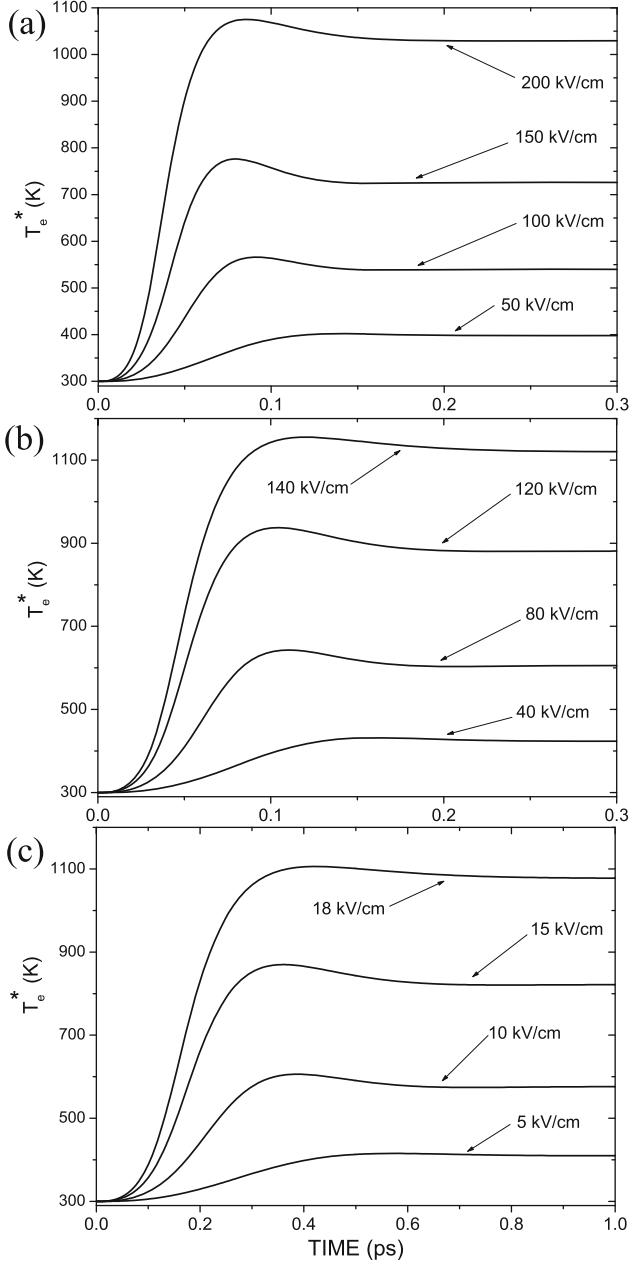

**Figure 1.** Evolution of electron non-equilibrium temperature for III nitrides, for several values of electric field intensity: (**a**) AlN, (**b**) GaN and (**c**) InN.

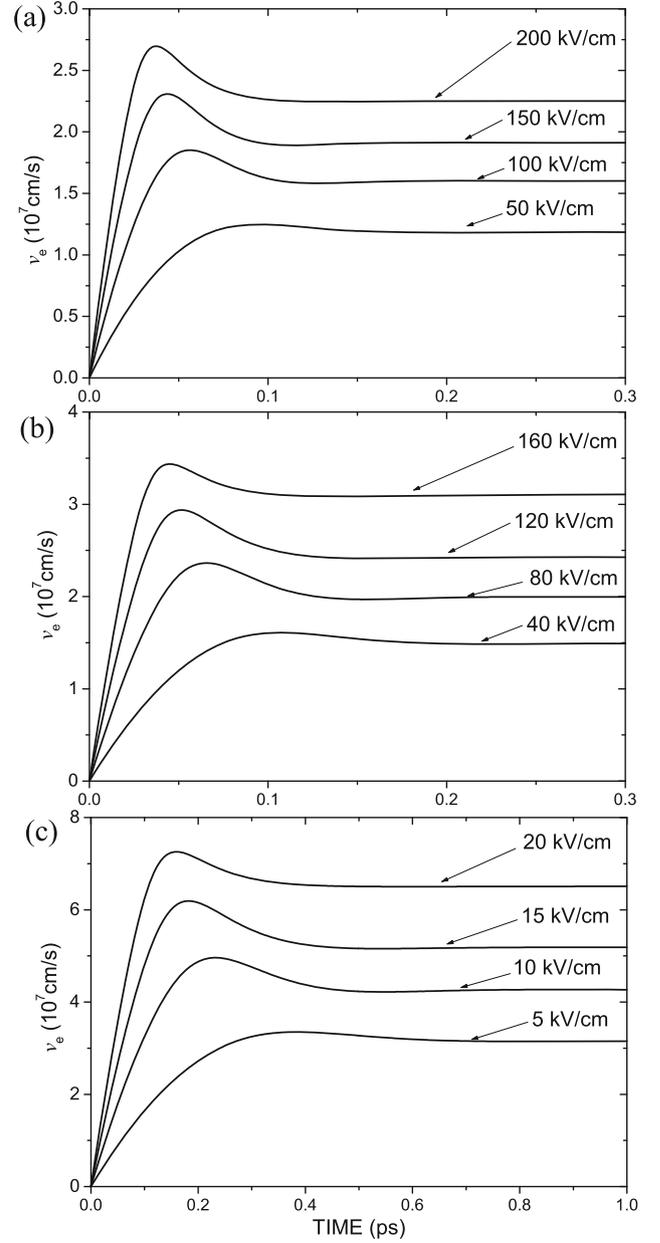

**Figure 2.** Evolution of drift velocity of electrons for III nitrides for several values of electric field intensity: (**a**) AlN, (**b**) GaN and (**c**) InN.

Figure 5 shows the quantity

$$x(t) = \frac{m_e^* v_e^2(t)}{k_B T_e^*(t)},$$

which is the ratio between the kinetic energy associated with the drift velocity in the presence of the electric field and the thermal kinetic energy: the peaks indicate the presence of a more pronounced overshoot in velocity than in quasitemperature. Furthermore, for the range of electric fields shown, $x$ in the steady state is less than one, and appears to remain so for higher fields, implying that the thermal kinetic energy is always higher than the kinetic energy due to the drifting movement.

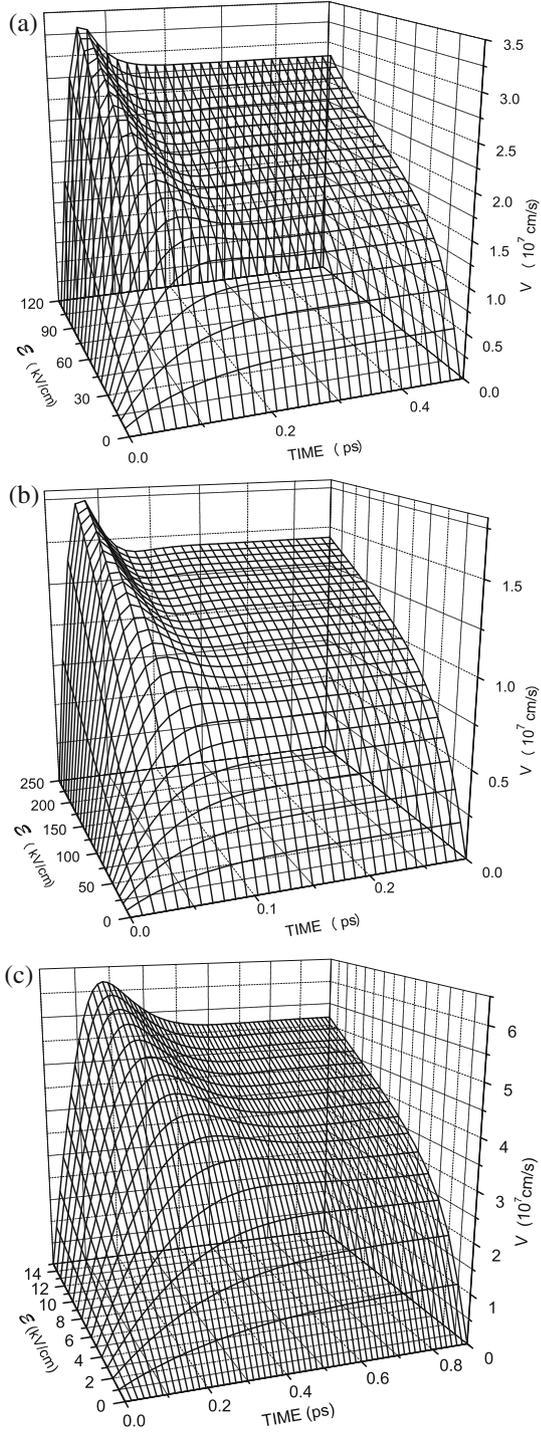

**Figure 3.** Time evolution of electron drift velocity: (**a**) GaN, (**b**) AlN and (**c**) InN.

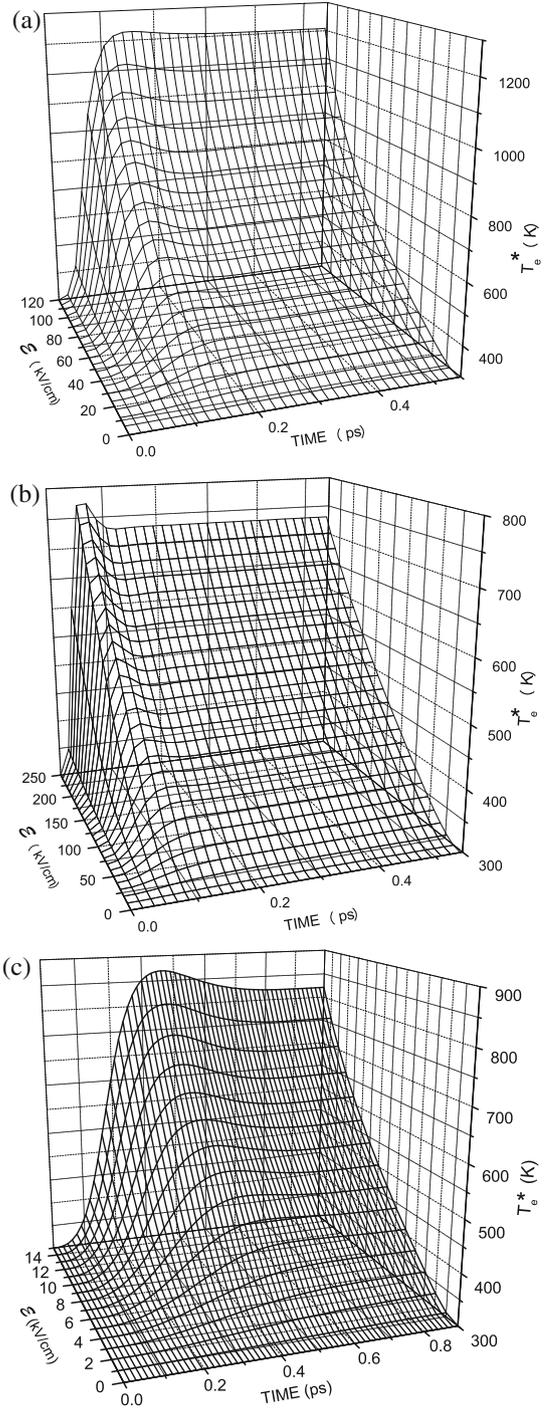

**Figure 4.** Time evolution of electron non-equilibrium temperature: (**a**) GaN, (**b**) AlN and (**c**) InN.

Moreover, figure 6 allows us to compare the momentum and energy relaxation times, $\tau_{P_e}(t)$ and $\tau_{E_e}(t)$, defined as

$$\tau_{P_e}(t) = \frac{P_e}{J_{P_e}^{(2)}(t)} , \tag{48}$$

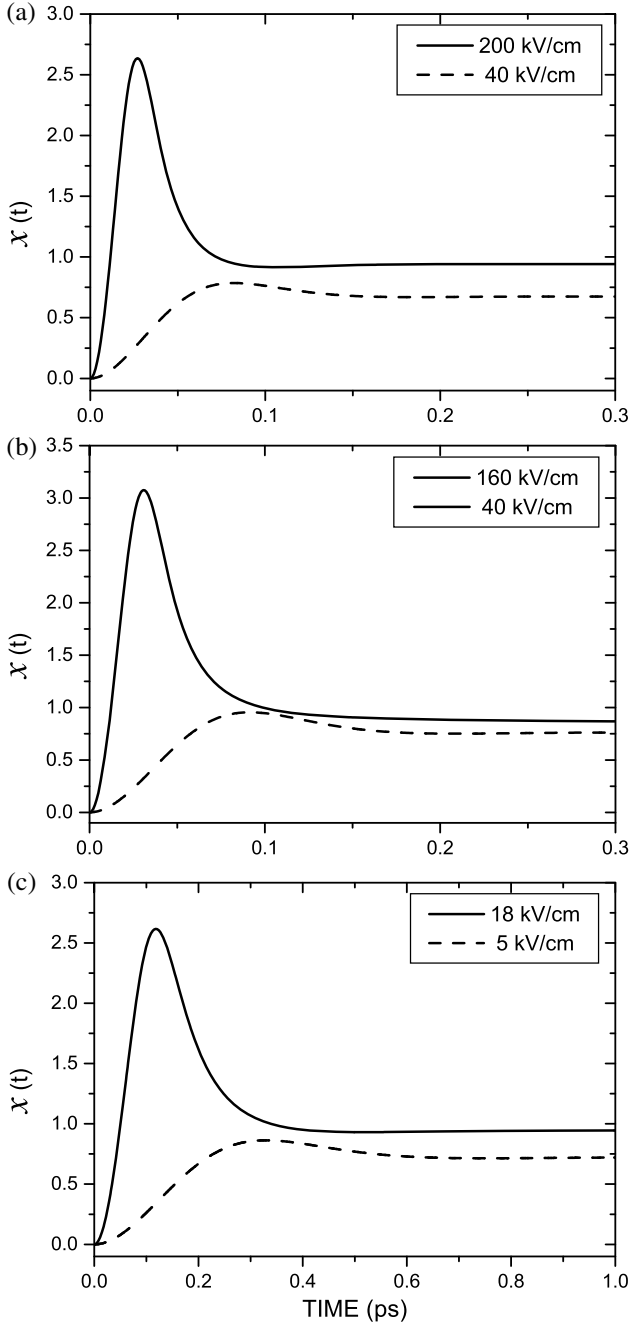

**Figure 5.** The ratio between kinetic energy due to drift velocity in the electric field and the kinetic thermal energy: (**a**) AlN, (**b**) GaN and (**c**) InN.

$$\tau_{E_e}(t) = \frac{E_e(t) - E_e^{\text{equil}}}{J_{E_e}^{(2)}(t)} \ , \tag{49}$$

where $J^{(2)}$ represents the collision operators in eqs (16) and (18). The relaxation times for InN (not shown in figure 6) are similar.

A similar study on p-doped III nitrides (similar to the results shown in figures 1–6) is available, and the results can be found in [56,64].

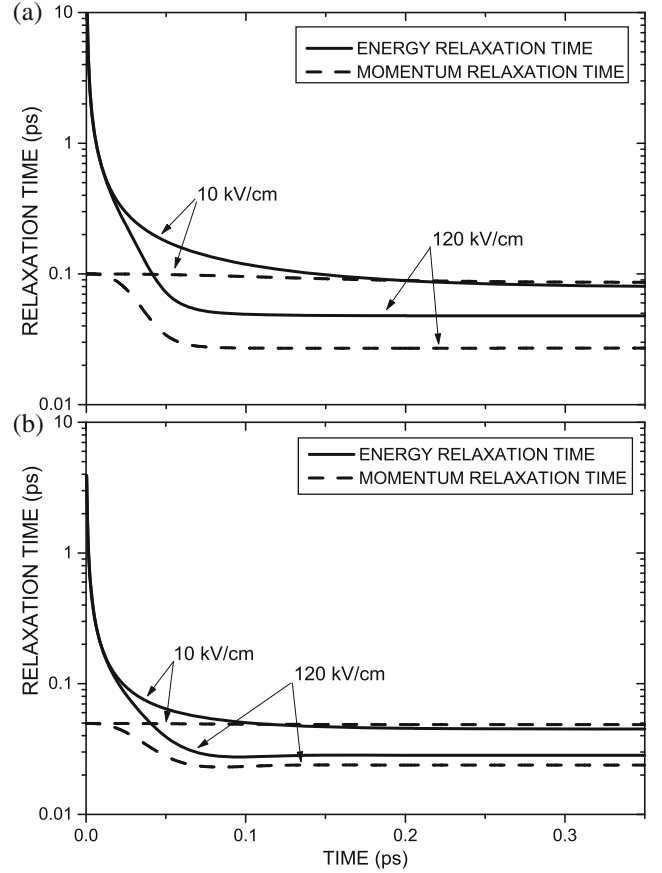

**Figure 6.** Comparison of energy and momentum relaxation times for two values of electric field intensity: (**a**) GaN and (**b**) AlN.

Recall that GaN is a wide-gap semiconductor that usually crystallises into a wurtzite (WZ) lattice (also known as hexagonal or $\alpha$-GaN). However, under certain conditions, zincblende (ZB) GaN (sometimes referred to as cubic or $\beta$-GaN) can be grown on zincblende substrates. Then, there are divergences in relation to the values of some basic parameters of GaN [120,121]. Seeking a better understanding of the influence of the electron effective mass $m_e^*$ on the electron drift velocity in n-doped GaN (WZ), figure 7 shows the time evolution of the electron drift velocity towards steady state in n-doped GaN (WZ) for four different values of electron effective mass [103,122–124] in the presence of an electric field of intensity 50 kV/cm. It was verified that after a transient time of the order of 200 fs to 300 fs, a steady state is attained, indicating that the electron drift velocity depends critically on the value of the electron effective mass. When examining figure 7, we note that the lower the electron effective mass is, the higher will be the velocity overshoot.

Figure 8 shows the time evolution of the electron drift velocity towards steady state in n-doped AlN (WZ)

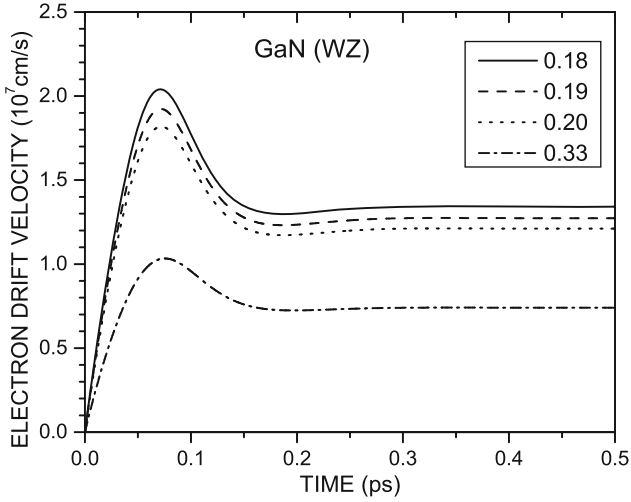

**Figure 7.** Evolution of electron drift velocity in n-doped GaN (WZ), for different values of electron effective mass: $m_e^* = 0.18m_0$, ref. [122]; $m_e^* = 0.19m_0$, ref. [103]; $m_e^* = 0.20m_0$, ref. [123]; $m_e^* = 0.33m_0$, ref. [124], where $m_0$ is the electron rest mass.

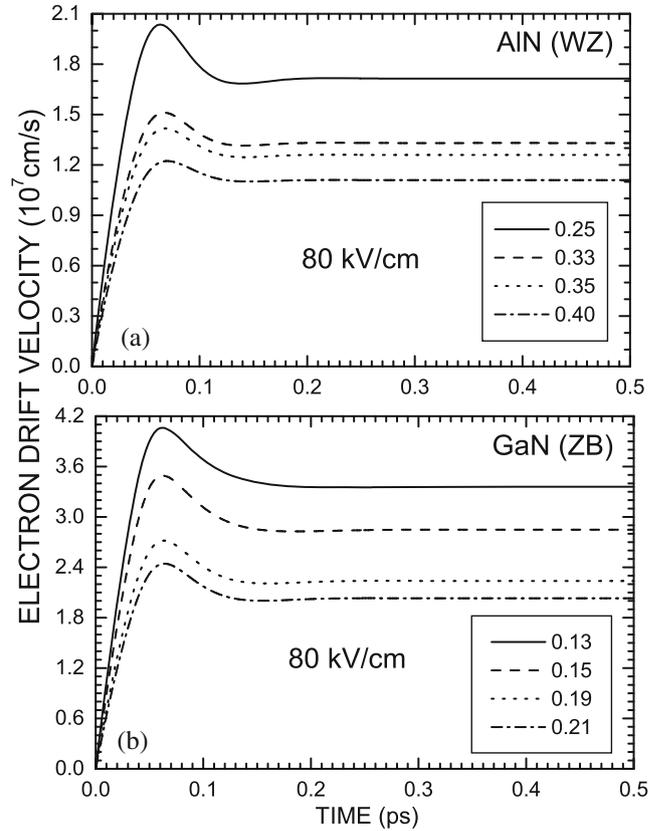

**Figure 8.** Time evolution of electron drift velocity for different values of electron effective mass: (**a**) AlN (WZ): $m_e^* = 0.25m_0$, ref. [125]; $m_e^* = 0.33m_0$, ref. [126]; $m_e^* = 0.35m_0$, ref. [103]; $m_e^* = 0.40m_0$, ref. [124]; (**b**) GaN (ZB): $m_e^* = 0.13m_0$, ref. [127]; $m_e^* = 0.15m_0$, ref. [128]; $m_e^* = 0.19m_0$, ref. [103]; $m_e^* = 0.21m_0$, ref. [105], where $m_0$ is the electron rest mass.

(figure 8a) and Ga (ZB) (figure 8b) for four different values of electron effective mass [103,105,124–128] in the presence of an electric field of intensity 80 kV/cm. It is verified that the steady state is attained after a transient time of the order of approximately 0.25 ps. Examining figure 8, we observe that the lower the electron effective mass is, the higher will be the velocity overshoot. For this electric field with an intensity of 80 kV/cm, the electron drift velocity (in the steady state) in GaN (ZB) for $m_e^* = 0.13m_0$ is $3.3 \times 10^7$ s and for $m_e^* = 0.21m_0$ is $2.0 \times 10^7$ cm/s. That is, the velocity differs by approximately 39%. For AlN (WZ), the electron drift velocity (in the steady state) for $m_e^* = 0.25m_0$ is $1.7 \times 10^7$ cm/s and for $m_e^* = 0.40m_0$, it is $1.1 \times 10^7$ cm/s, resulting in a difference of approximately 35%.

The transient transport regime in the III nitrides follows a sub-picosecond scale during which an overshoot in the electron velocity occurs for sufficiently high electric fields. The onset of the electron velocity overshoot in wurtzite InN was shown by O'Leary *et al* [112] to occur for electric fields higher than 22.5 kV/cm, and the same value induced negative differential resistivity (NDR) [19,112,129–131]. InN was initially reported to have an energy gap of 1.8 to 2.0 eV [35,36]. However, since 2001, values from 0.67 eV (low temperature) to 0.63 eV (room temperature) have been reported based on the optical properties of high-quality hexagonal InN epitaxial layers with low background electron densities grown by molecular beam epitaxy (MBE) [29,30,34]. In addition, while initial experimental results suggested an electron effective mass ($m_e^*$) of $0.11m_0$ [36], more recent experimental measurements have instead suggested an electron effective mass of $0.045m_0$ [33] for this material (here $m_0$ is the free electron mass).

Figures 9 and 10 respectively show the time evolution of the electron drift velocity and electron's non-equilibrium temperature towards steady state in n-doped InN for five different values of electron effective mass in the presence of an electric field of intensity 10 kV/cm. When examining figure 9, note that the lower the electron effective mass is, the higher will be the velocity overshoot. The overshoot effect follows if during the time evolution of the macroscopic state of the system, under the action of an electric field, the electron relaxation rate of momentum is larger than the electron relaxation rate of energy. It is verified that the electron drift velocity depends critically on the value of the electron effective mass. For an electric field of intensity 10 kV/cm, and for $m_e^* = 0.35m_0$, the electron drift velocity is $5.4 \times 10^7$ cm/s (in the steady state) and $6.5 \times 10^7$ cm/s (at the peak), indicating a difference of approximately 16.9%. For $m_e^* = 0.055m_0$, the electron drift velocity is

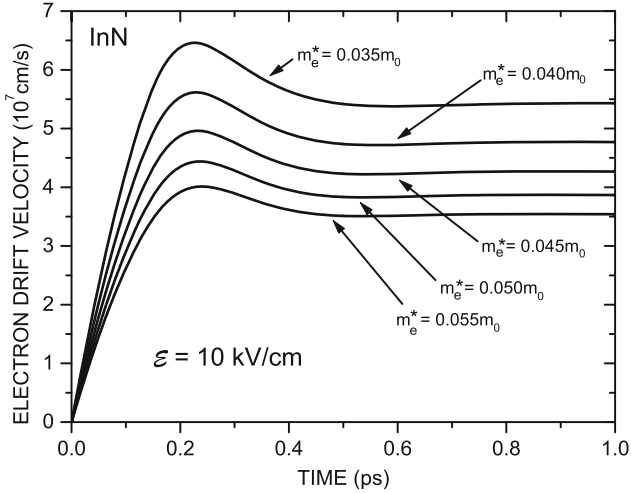

**Figure 9.** Time evolution of electron drift velocity in n-doped InN, for different values of electron effective mass.

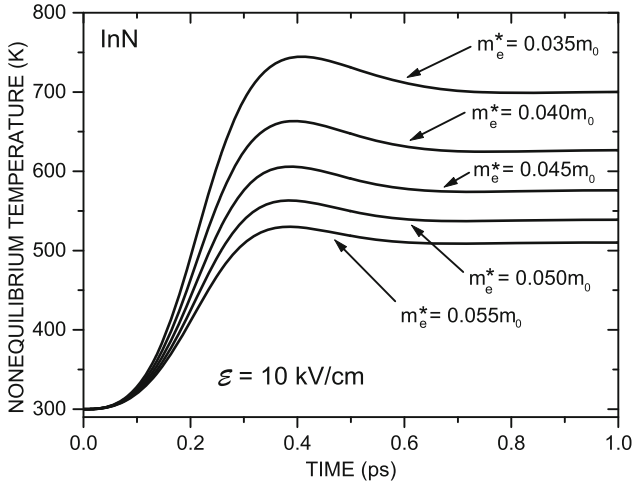

**Figure 10.** Time evolution of electron non-equilibrium temperature in n-doped InN, for different values of electron effective mass.

$3.5 \times 10^7\,\text{cm/s}$ (in the steady state) and $4.0 \times 10^7\,\text{cm/s}$ (at the peak), resulting in a difference of approximately 12.5%.

Figures 7–10 show the importance of accurately determining the value of electron effective mass for performing theoretical calculations and simulations of electronic transport in III nitrides (GaN, AlN and InN) because the results obtained depend critically on the value assumed for the electron effective mass.

### 3.2 Steady state

Consider the steady state, which occurs very rapidly (in 100 fs time scale), which can be understood on the basis of the action of the intense Fröhlich interaction in these strong polar semiconductors, with the rate of transfer of energy from carriers to LO phonons rapidly equalising the rate of energy being pumped from the external field, even at high fields. In the steady state, $\mathrm{d}E_e(t)/\mathrm{d}t = 0$, $\mathrm{d}\mathbf{P}_e(t)/\mathrm{d}t = 0$, $\mathrm{d}E_{\mathrm{LO}}(t)/\mathrm{d}t = 0$ and $\mathrm{d}E_{\mathrm{AC}}(t)/\mathrm{d}t = 0$ (see eqs (16)–(20)).

Figure 11 shows the dependence in the steady state of the non-equilibrium temperature of the carriers on the electric field strength $\mathcal{E}$. A near-parabolic dependence can be observed, more precisely of the form $A + B\mathcal{E} + C\mathcal{E}^2$, and note that the figures have end points because beyond those points, it is necessary to introduce a more detailed band structure (intervalley scattering begins to become relevant), superseding the parabolic band approximation that was used.

Figure 12 describes the dependence in the steady state of the electron drift velocity on the electric field for three electron concentrations when the lattice temperature is 300 K. The relationship is not Ohmic, with a near-Ohmic law occurring only at low fields. For larger intensities of the electric field, nonlinear transport follows, and the current increases less markedly with $\mathcal{E}$, that is, the angular coefficient $\mathrm{d}v_e/\mathrm{d}\mathcal{E}$ continues to decrease with increasing $\mathcal{E}$. The regions exhibiting Ohmic and non-Ohmic behaviour when considering the electron drift velocity dependence on the electric field strength are shown in figure 13 when the electron concentration is $10^{17}\,\text{cm}^{-3}$. At low fields, an Ohmic region is present. However, it is limited to fields of roughly 15 kV/cm, 12 kV/cm and 1.5 kV/cm for AlN, GaN and InN respectively. Beyond these values, a departure from Ohmic behaviour takes place (note the break in scale in the horizontal axis).

Moreover, the mobility $\mathcal{M}$ of the carriers, given by

$$\mathcal{M} = \frac{|\mathbf{v}_e|}{|\mathcal{E}|}\,, \tag{50}$$

is shown in figure 14 in terms of the electric field intensity for an electron concentration of $10^{17}\,\text{cm}^{-3}$ and a lattice temperature of 300 K. Note that the higher mobility corresponds to InN, which can be attributed to the fact that the electrons have a lower effective mass (see table 1) than in GaN and AlN: $\mathcal{M}_{\mathrm{InN}} > \mathcal{M}_{\mathrm{GaN}} > \mathcal{M}_{\mathrm{AlN}}$.

Recall that figures 11 and 12 show the dependence on the electric field, and in the steady state, of the non-equilibrium temperature and the drift velocity for the three n-type III-nitrides with $n = 10^{17}\,\text{cm}^{-3}$, and a lattice temperature of 300 K. The non-equilibrium temperature follows a nearly quadratic (parabolic) law with respect to $\mathcal{E}$, while the drift velocity, after a near-Ohmic regime (linearity with $\mathcal{E}$) at low fields, increases with a decreasing slope (implying a decreasing differential conductivity) as the intensity of the field $\mathcal{E}$ increases. These effects are better observed in figure 15, where the dependence of the mobility on the electric field is illustrated, together with that of the differential mobility $\mathrm{d}v_e/\mathrm{d}\mathcal{E}$. It follows that after displaying a near constant value at low fields ($\lesssim 5$ kV/cm), corresponding to a near-Ohmic regime, both quantities continue to decrease with increasing intensity of the electric field, with the differential mobility being lower than mobility. This is because the momentum relaxation time continues to decrease as the intensity $\mathcal{E}$ of the electric field increases. Such dependence follows basically from the dependence of the non-equilibrium temperature $T_e^*$, where the momentum relaxation time decreases as the latter increases, that is, with increasing thermal agitation [59].

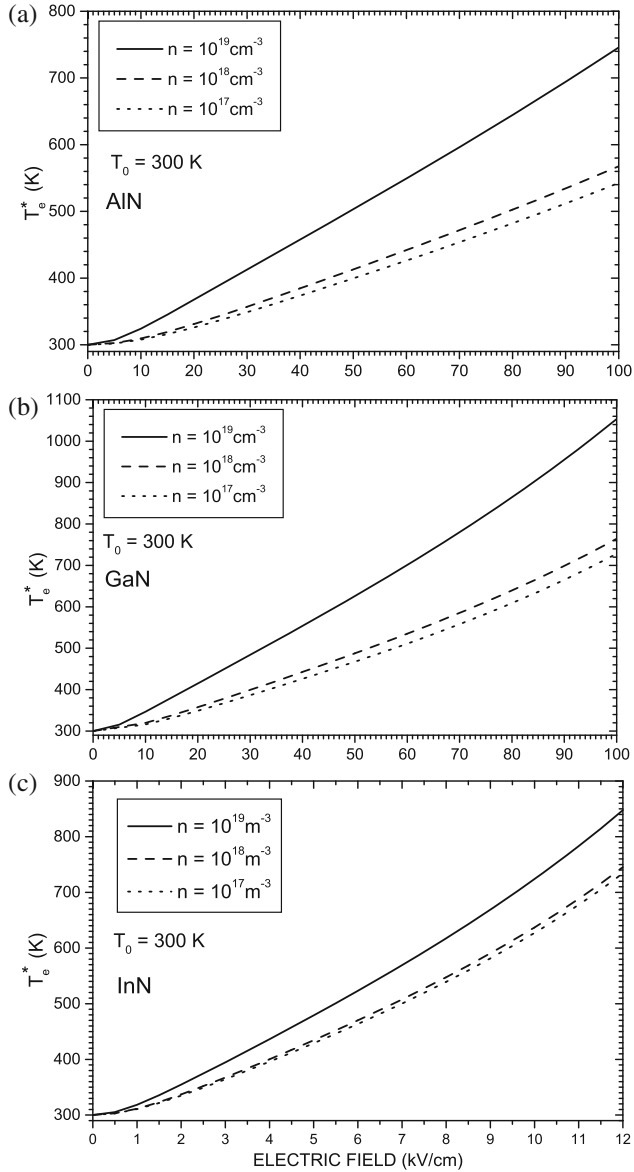

**Figure 11.** Dependence of electron non-equilibrium temperature $T_e^*$ on electric field intensity for three electron concentrations, in steady state of the semiconductors: (**a**) AlN, (**b**) GaN and (**c**) InN. Lattice temperature $T_0 = 300$ K.

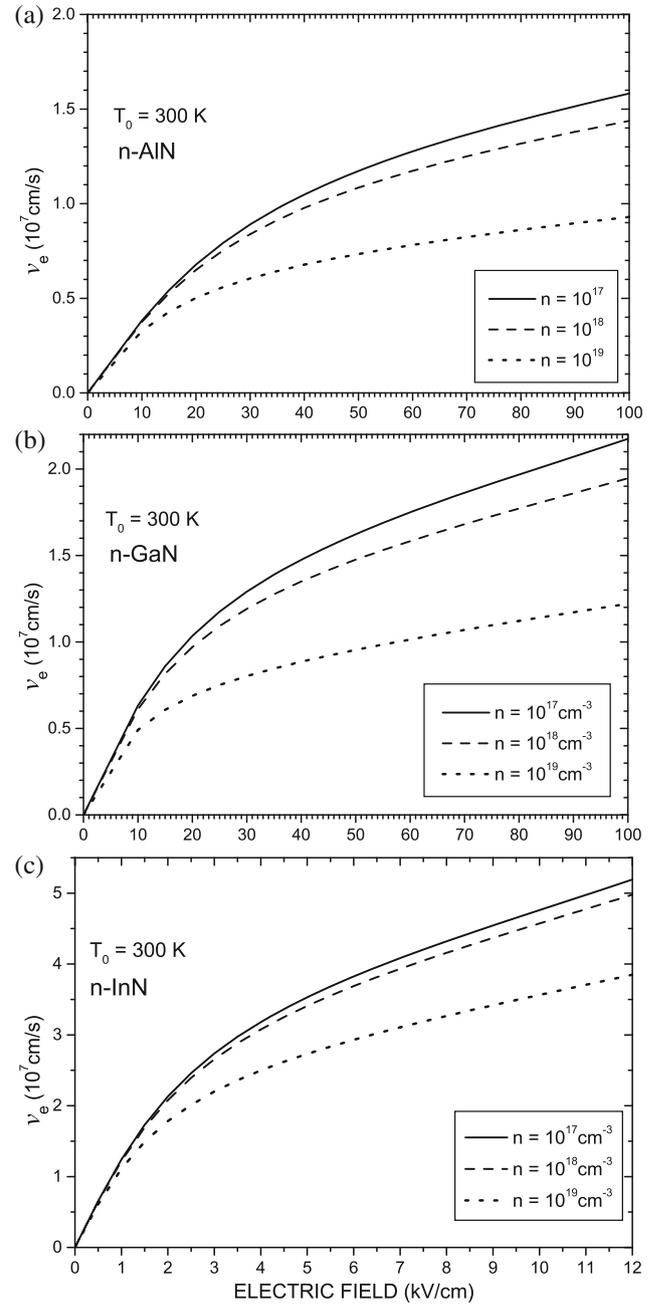

**Figure 12.** Dependence of electron drift velocity $v_e$ on electric field intensity for three electron concentrations, in steady state of the semiconductors: (**a**) AlN, (**b**) GaN and (**c**) InN. Lattice temperature $T_0 = 300$ K.

The linear momentum density can be related to the drift velocity $v_e(t)$ (along the $x$-axis, i.e., parallel to the electric field) by the relationship

$$P_e = N m_e^* v_e , \qquad (51)$$

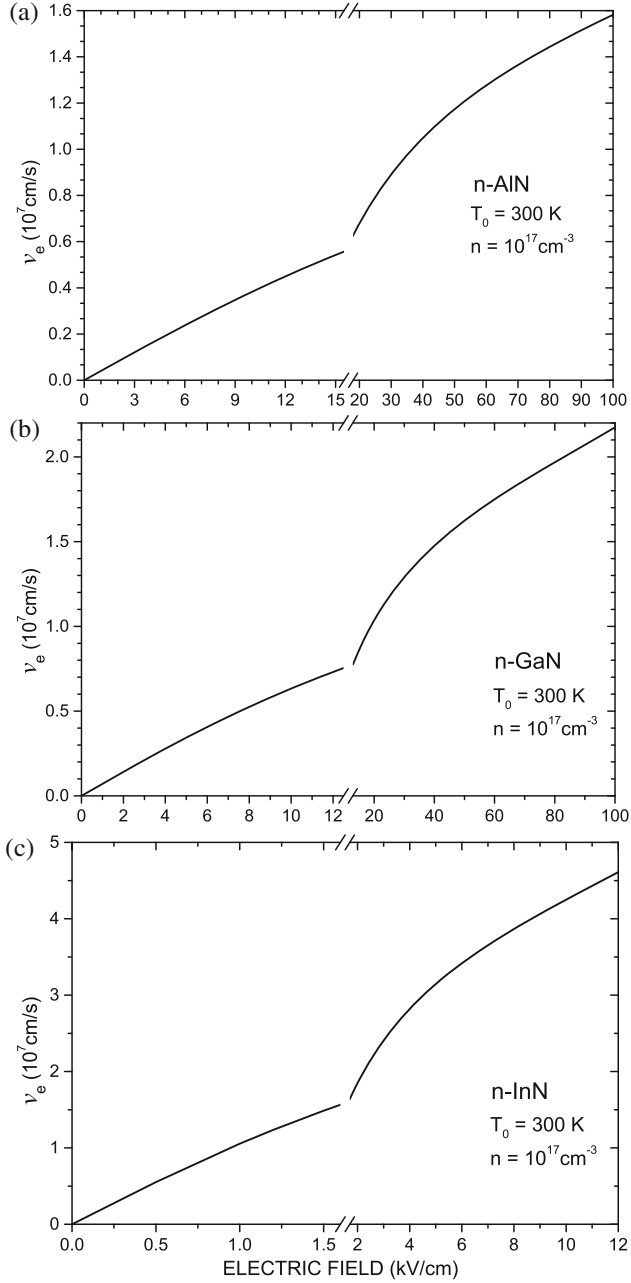

**Figure 13.** Steady-state drift velocity of the carriers as a function of electric field: (**a**) AlN, (**b**) GaN and (**c**) InN.

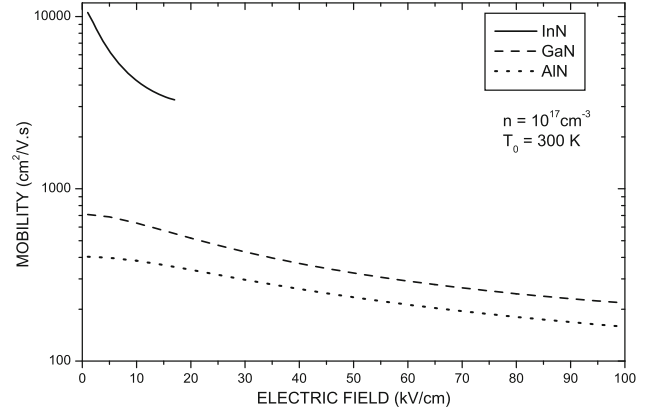

**Figure 14.** Dependence of carrier mobility on electric field intensity, in steady state of three semiconductors indicated in the upper-right inset. Lattice temperature is 300 K and electron concentration is $10^{17}\,\mathrm{cm}^{-3}$.

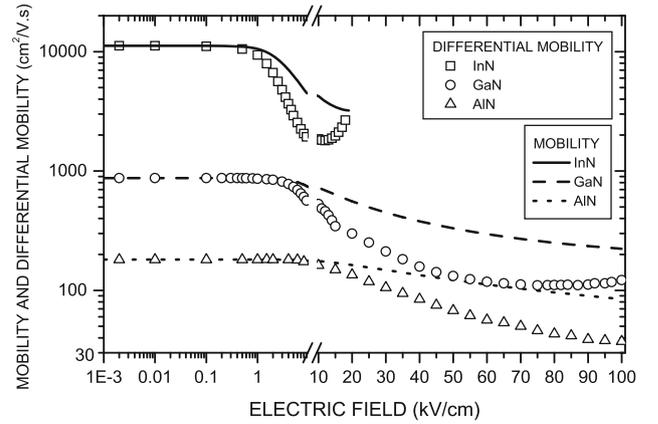

**Figure 15.** Mobility and the differential mobility as a function of electric field intensity (with a change from a logarithmic to a linear scale at 10 kV/cm).

where $m_e^*$ is the effective mass of the electron and is related to the current $I_e$ by the expression

$$I_e(t) = -nev_e \,, \tag{52}$$

which flows in the direction of the electric field. Moreover, we define the momentum relaxation time associated with the scattering of carriers by phonons as

$$\tau_{P_e}(t) = -n \frac{m_e^* v_e}{J_{P_e}^{(2)} + J_{P_e,\mathrm{imp}}^{(2)}} \,, \tag{53}$$

where once we take into account that the collision operator $J_{P_e}^{(2)}$ is composed of three contributions consisting of scattering by the optical (or Fröhlich) interaction among LO phonons, deformation potential of LO and AC phonons, and through the piezoelectric potential of AC phonons, we have a Mathiessen-like rule of the form

$$\frac{1}{\tau_{P_e}} = \frac{1}{\tau_{\mathrm{PO}}} + \frac{1}{\tau_{\mathrm{AD}}} + \frac{1}{\tau_{\mathrm{PZ}}} + \frac{1}{\tau_{\mathrm{imp}}} \,. \tag{54}$$

In eq. (54), PO refers to the polar-optic interaction, AD represents the acoustic deformation potential, PZ to acoustic piezoelectric potential, and 'imp' is the effect of impurities, which can be expressed as individual relaxation times

$$\tau_i = \frac{nm_e^* v_e}{J_i^{(2)}} \,, \tag{55}$$

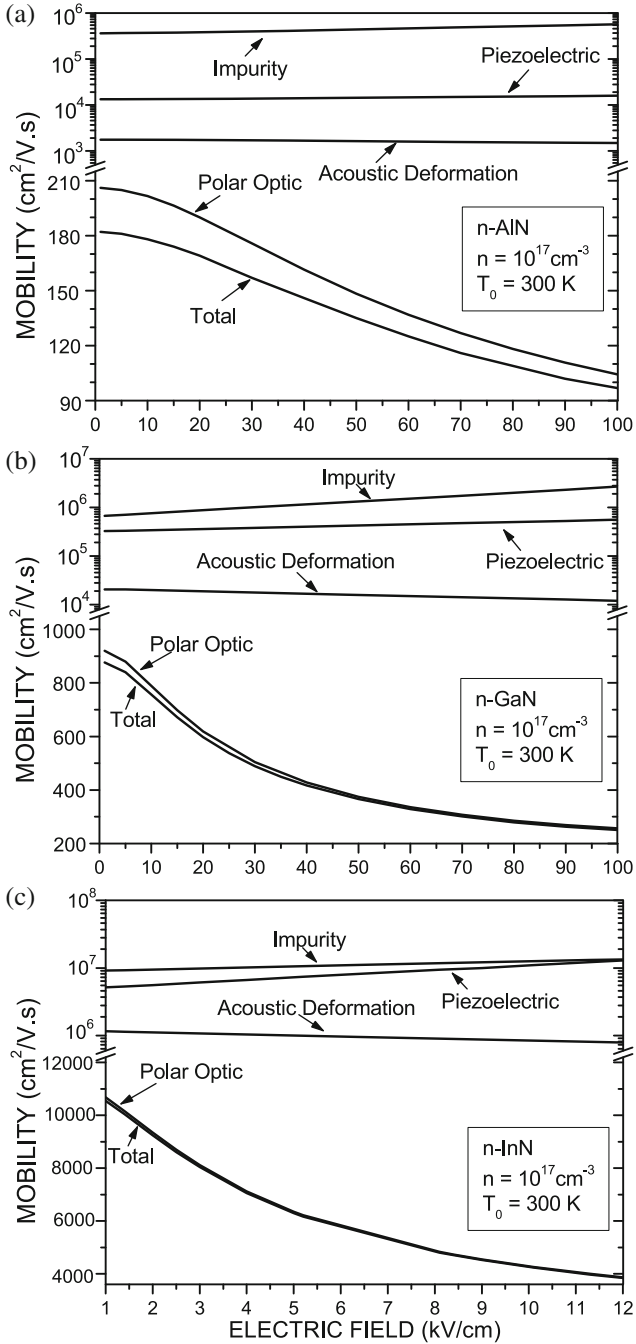

**Figure 16.** Electric field dependence of different contributions to electron mobility: (**a**) AlN, (**b**) GaN and (**c**) InN. Lattice temperature $T_0 = 300$ K and $n = 10^{17}$ cm$^{-3}$.

where $J_i^{(2)}$ is the contribution to $J_{P_e}^{(2)}$ in eq. (15) for $i$ stands for PO, AD, or PZ and $i =$ 'imp' for $J_{imp}^{(2)}$.

Consequently, the mobility in the steady state, namely,

$$\mathcal{M}_e = \frac{v_e}{\mathcal{E}} = \frac{e}{m_e^*}\, \tau_{P_e}\,, \tag{56}$$

(once in the steady state, according to eq. (15), $J_{P_e}^{(2)} + J_{P_e,\mathrm{imp}}^{(2)} = ne\mathcal{E}$) becomes a composition of four contributions, according to the rule

$$\frac{1}{\mathcal{M}_e} = \frac{1}{\mathcal{M}_{\mathrm{PO}}} + \frac{1}{\mathcal{M}_{\mathrm{AD}}} + \frac{1}{\mathcal{M}_{\mathrm{PZ}}} + \frac{1}{\mathcal{M}_{\mathrm{imp}}}\,. \tag{57}$$

The numerical results are shown in figure 16 for the three III nitrides. Figure 16 provides a comparison of the four contributions to the mobility (see eq. (57)). Note that this indicates a valid Mathiessen rule, such that the much lower $\mathcal{M}_{\mathrm{PO}}$ essentially determines the value of the entire mobility. This clearly indicates that the polar-optic (Fröhlich) interaction is strong in these compounds, and therefore produces a very short relaxation time compared to the other scattering mechanisms. In addition, note that under the conditions of the calculations, scattering caused by the ionised impurities is negligible in comparison to the others. Furthermore, note that the break in scale in the vertical axis initially follows a linear scale before the break, and follows a log scale after the break.

Figures 17a–17c show the mobility in steady state in terms of the concentration for different values of electric field strength at a lattice temperature of 300 K. The larger mobility corresponds to InN, which can be ascribed to the fact that as pointed out previously, the electrons have a lower effective mass in InN than in AlN and GaN. The mobility is proportional to the relaxation time, which consists of the contributions due to scattering of phonons and impurities, but the latter is orders of magnitude greater than the former, such that the mobility is approximately proportional to the momentum relaxation time. This depends on the macrostate of the system, that is, on the quasitemperature $T_e^*$ and drift velocity $v_e$ as well as on the concentration of carriers $n$ and the quasitemperature of LO phonons, $T_{\mathrm{LO}}^*$. Inspection of these curves shows that the dependence of mobility on concentration is nearly constant up to $n \sim 5 \times 10^{17}$ cm$^{-3}$. For $n$ larger than $\sim 10^{18}$ cm$^{-3}$, the mobility continues to decrease. These results indicate the presence of two distinct regimes, one at low concentrations (up to $10^{18}$ cm$^{-3}$) and the other at higher concentrations (greater than $10^{19}$ cm$^{-3}$).

Figure 18 shows mobility normalised to its value at $n = 10^{17}$ cm$^{-3}$ vs. concentration under an electric field of 40 kV/cm, where the continuous line is the result of the calculations (see figure 17b) and the dots are the values obtained from the best fitting using the expression

$$\mathcal{M}^* = \frac{\mathcal{M}}{\mathcal{M}_0} = A_0 + A_1 \mathrm{e}^{-n/n_1} + A_2 \mathrm{e}^{-n/n_2}\,, \tag{58}$$

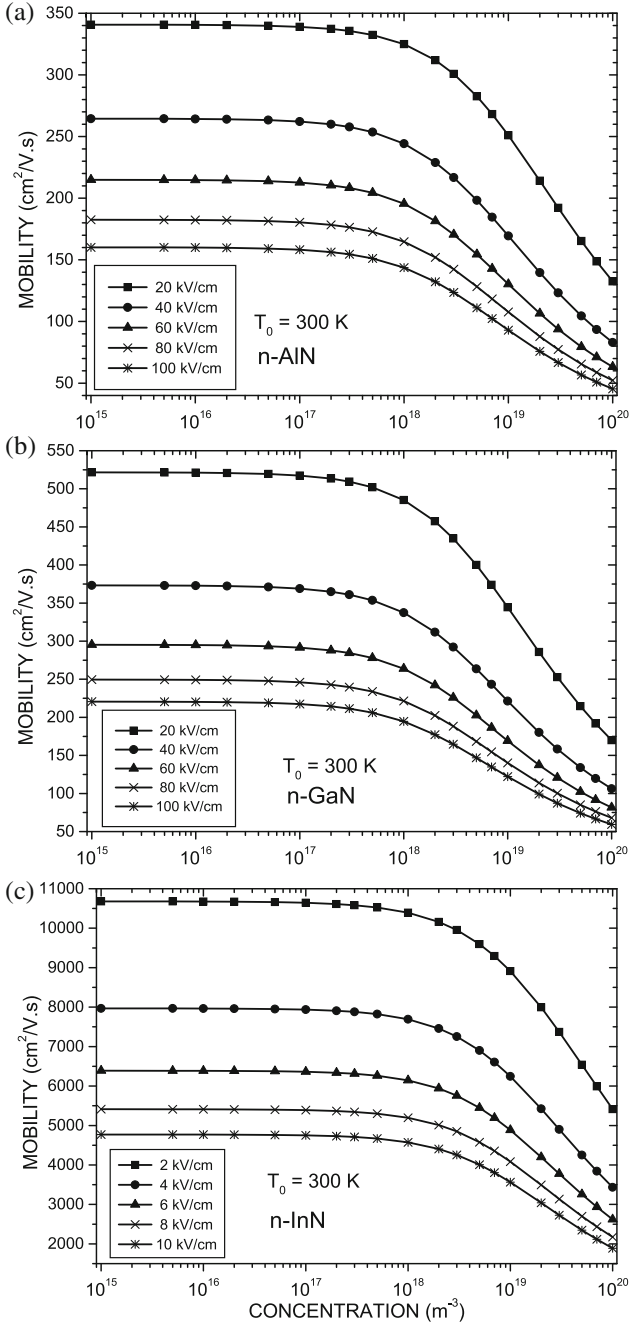

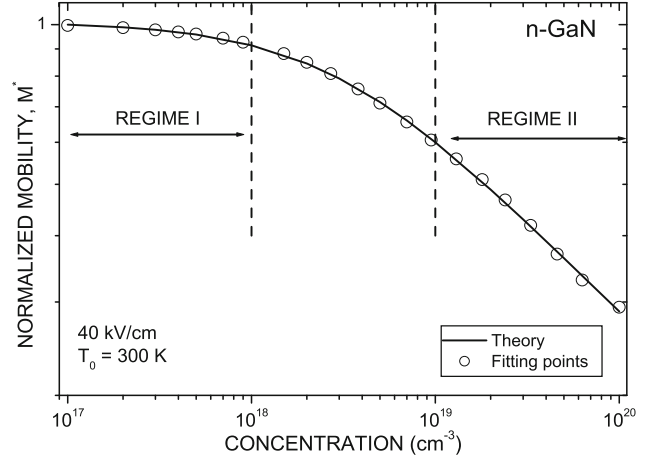

**Figure 18.** Dependence of normalised mobility ($\mathcal{M}^* = \mathcal{M}/\mathcal{M}_0$) on carrier concentration in n-GaN for the case of a field of 40 kV/cm and $T_0 = 300$ K. Full line was obtained from the theory (see figure 17b) and the open circles are values obtained from best fitting according to eq. (58).

**Figure 17.** Dependence on carrier concentration, for several values of electric field strength, of the carrier mobility in steady state: (**a**) AlN, (**b**) GaN and (**c**) InN.

where $A_0 \simeq 0.28$, $A_1 \simeq 0.42$, $A_2 \simeq 0.31$, $n_1 \simeq 3.03 \times 10^{19}$ cm$^{-3}$, $n_2 \simeq 3.68 \times 10^{18}$ cm$^{-3}$. The presence of the two regimes is clearly evidenced: at lower concentration, the contribution with index 1 predominates, while at higher concentration the contribution with index 2 predominates (see eq. (58)).

Moreover, it is interesting to note that in each of the aforementioned regimes, mobility vs. concentration approximately follows a fractional power law, that is

$$\mathcal{M}^* \simeq a \cdot n^b , \qquad (59)$$

where in regime I, that is, for $n < 10^{18}$ cm$^{-3}$, $a \simeq 4.42$ and $b \simeq -0.04$, and in regime II, that is, for $n > 10^{19}$ cm$^{-3}$, $a \simeq 1.12 \times 10^6$ and $b \simeq -0.33$.

Regarding these results, many empirical analysis suggest that a non-integer power-law behaviour of some quantities occurs frequently in nature. Many researchers have attempted to relate this fact to the complex behaviour observed in the system, speculating that it arises out of chaotic or fractal properties. However, these non-integer power laws may be purely apparent or simply spurious [132], as seems to be true in the present case.

The influence of temperature of the acoustic phonons – assumed to be constantly in near equilibrium with the thermal reservoir – on quasitemperatures (non-equilibrium temperatures) and mobility is shown in figures 19 and 20, respectively, for a concentration of $10^{17}$ cm$^{-3}$ and an electric field of 40 kV/cm. The relaxation time of the LO-phonon populations resulting from the anharmonic interactions is typically in the range of a few picoseconds [133]. For the calculations, we used 5 ps (the results are weakly dependent on $\tau_{an}$, and this relaxation time can fall within the range of 1–10 ps).

It can be observed that the quasitemperature of the LO-phonons practically coincides with the AC-phonons (and reservoir) temperature, that is, both subsystems are in thermal equilibrium. Moreover, as $T_0$ increases, the difference between $T_e^*$ and $T_{LO}^*$ continues to decrease because the exchange of energy becomes more efficient between both subsystems. The carriers' non-equilibrium temperature is, for each given value of the electric field strength, higher in GaN than in AlN. This is a consequence of the differences in the effective mass and in the Fröhlich coupling parameter $\alpha$:

$$\alpha = \sqrt{\frac{e^4 m_e^*}{2\hbar\omega_{\text{LO}}}} \left( \frac{1}{\varepsilon_\infty} - \frac{1}{\varepsilon_0} \right). \tag{60}$$

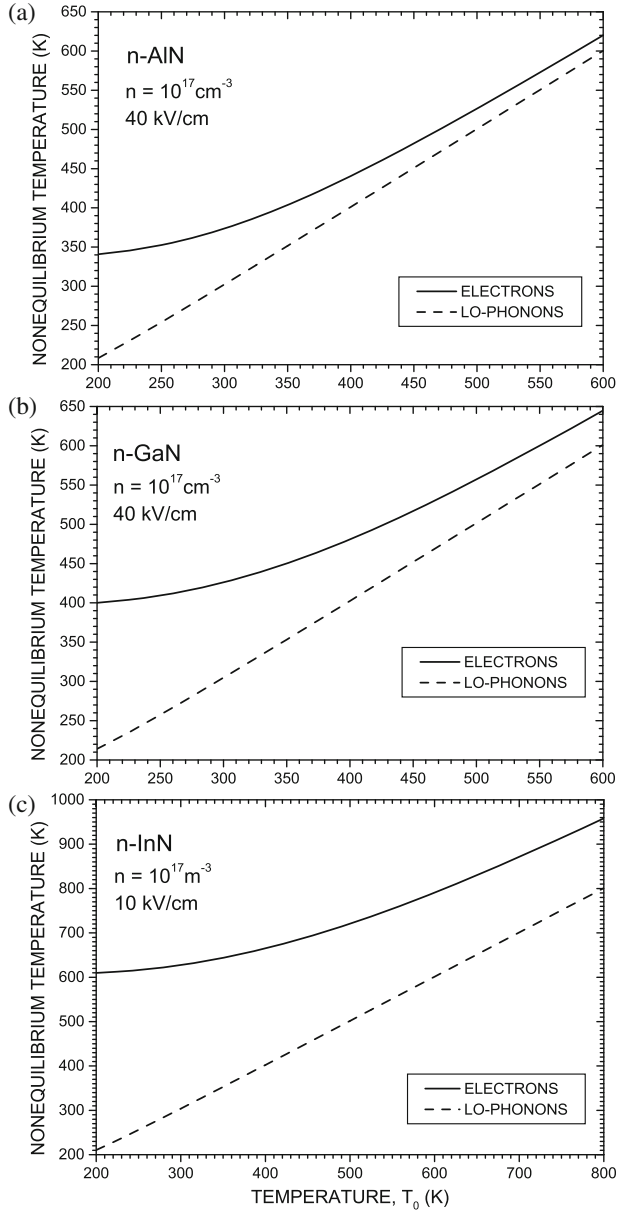

**Figure 19.** Dependence of non-equilibrium temperatures, when $n = 10^{17}\,\text{cm}^{-3}$, of electrons and LO-phonons on $T_0$ in (**a**) n-AlN, (**b**) n-GaN and (**c**) n-InN.

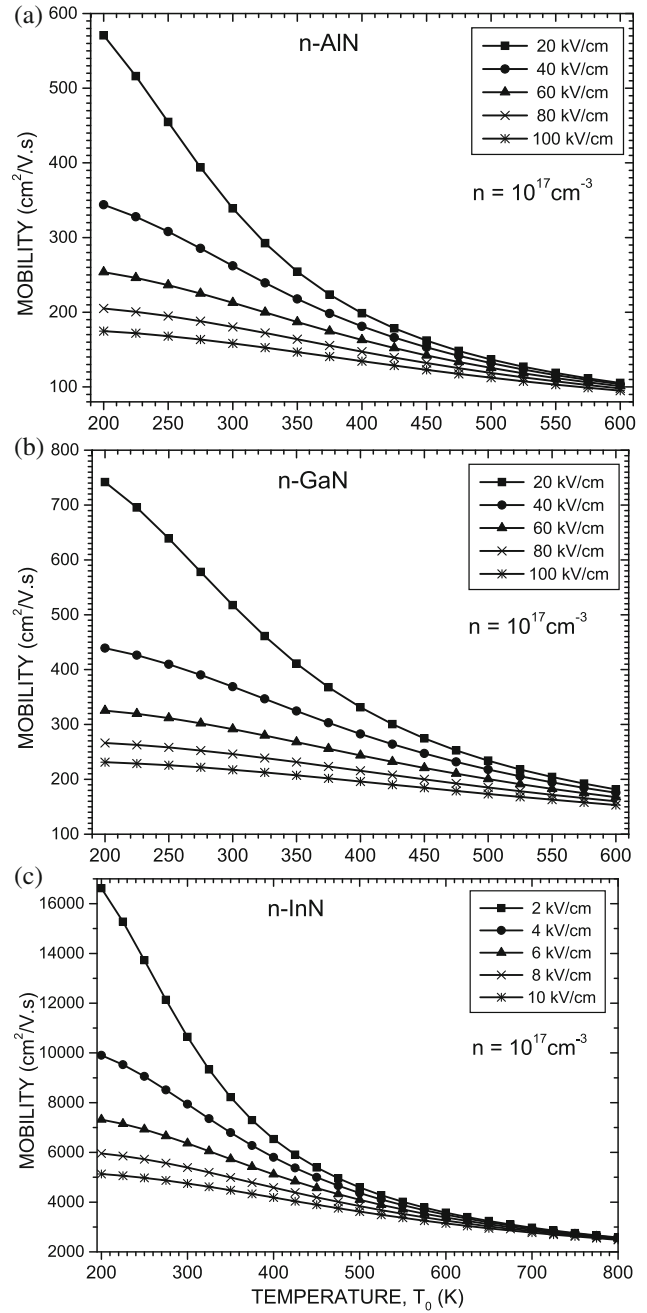

**Figure 20.** Dependence of carrier mobility, for several values of electric field strength with $n = 10^{17}\,\text{cm}^{-3}$, in steady state on $T_0$ in (**a**) n-AlN, (**b**) n-GaN and (**c**) n-InN.

The values for $m_e^*$ are given in the first line of table 1, and $\alpha = 0.42$ for GaN, 0.60 for AlN and 0.13 for InN.

This behaviour of non-equilibrium temperature of the carriers (recall that in this case, the quasitemperature of the LO phonons remains nearly equal to that of the thermal bath of acoustic phonons, $T_0$) is a determinant in the behaviour of mobility, as shown in figure 20. As $T_0$ increases, mobility continues to decrease in a similar manner for all values of the electric field strength. The mobility is lower when the field strength is higher. Moreover, with increasing thermal bath temperature, the mobility keeps decreasing with a tendency to be equal for all the field intensities considered. This is an expected result once the carrier quasitemperature approaches the LO phonon quasi-temperature (as previously observed in figure 19), which is nearly equal to $T_0$, and all the curves converge toward this value in thermal equilibrium. Therefore, the momentum relaxation time has similar values for all field intensities.

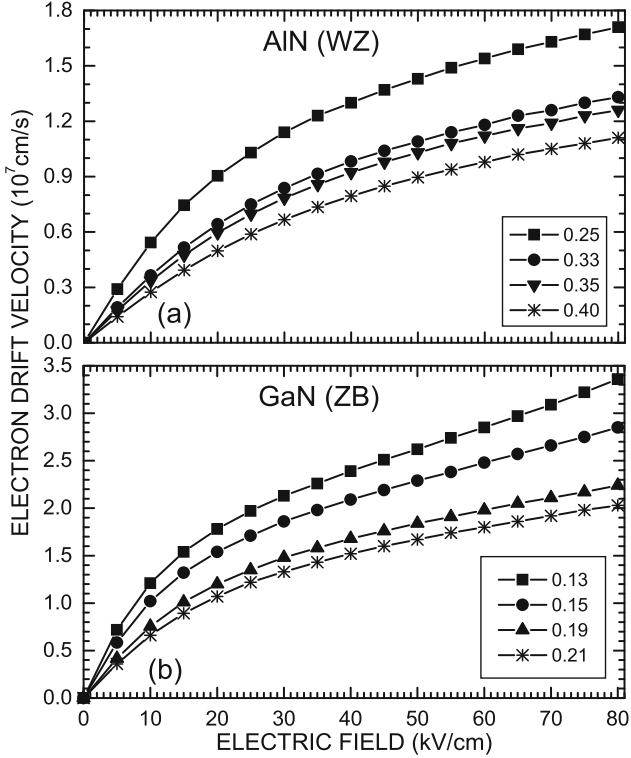

**Figure 21.** Dependence of electron drift velocity in steady state on electric field, for four different values of electron effective mass, in n-doped (**a**) AlN (WZ): $m_e^* = 0.25m_0$, ref. [125]; $m_e^* = 0.33m_0$, ref. [126]; $m_e^* = 0.35m_0$, ref. [103]; $m_e^* = 0.40m_0$, ref. [124] and (**b**) GaN (ZB): $m_e^* = 0.13m_0$, ref. [127]; $m_e^* = 0.15m_0$, ref. [128]; $m_e^* = 0.19m_0$, ref. [103]; $m_e^* = 0.21m_0$, ref. [105], where $m_0$ is the electron rest mass.

**Table 2.** Mobility of AlN (WZ) and GaN (ZB).

|  | $m_e^*$ ($m_0$) | Mobility ($c^2/V \cdot s$) |
|---|---|---|
| AlN (WZ) | 0.25 [125] | 599 |
|  | 0.33 [126] | 387 |
|  | 0.35 [103] | 352 |
|  | 0.40 [124] | 285 |
| GaN (ZB) | 0.13 [127] | 1560 |
|  | 0.15 [128] | 1252 |
|  | 0.19 [103] | 870 |
|  | 0.21 [105] | 744 |

Figures 21a and 21b show the dependence of electron drift velocity (in the steady state) on electric field strength, for four different values of electron effective mass when considering AlN (WZ) and GaN (ZB). At low electric fields, an Ohmic region is present, and a departure from the Ohmic behaviour takes place for larger fields. Table 2 shows the mobility (at low fields, i.e., the Ohmic region) for AlN (WZ) and GaN (ZB) for different values of electron effective mass. As mentioned previously, larger mobility values correspond to lower electron effective mass (of AlN [125] and GaN [127]).

Figure 22 shows the dependence of electron drift velocity in the steady state on the electron effective mass in n-doped GaN (WZ), for several values of electric field. For a fixed electric field of intensity 120 kV/cm, the electron drift velocity for $m_e^* = 0.18m_0$ is $1.93 \times 10^7$ cm/s and for $m_e^* = 0.33m_0$ it is $1.00 \times 10^7$ cm/s, which represents a difference of approximately 48%. For a fixed electric field of intensity 60 kV/cm, the electron drift velocity for $m_e^* = 0.18m_0$ is $1.43 \times 10^7$ cm/s and for $m_e^* = 0.33m_0$ it is $0.77 \times 10^7$ cm/s, resulting in a difference of approximately 46%. In general, this difference under different electric fields is approximately 45–55%.

### 3.3 *Comparisons with theoretical and experimental results*

Consider now the case of n-doped GaN, with $n = 10^{17}$ cm$^{-3}$ and four different temperatures $T_0$ of the thermal reservoir in contact with the sample, as considered in the Monte Carlo simulation conducted by Mansour *et al* [134], who reported that the values of the drift velocity in the steady state are functions of the electric field strength. This was found to be true in another study also [58]. Both results are compared in figure 23, and the results are in good agreement, with differences of less than 10%. Under high fields, the NESEF drift velocities are higher than those of the Monte Carlo simulation because a parabolic band model was used in the former, and at high fields (of the order of and higher than 100 kV/cm) scattering to side valleys in the band (with larger effective mass than the central valley) influences the current intensity. It can be observed that the drift velocity diminishes for each value of the field with increasing reservoir temperature. As expected, the relaxation of the energy beyond the equilibrium point of the system is less effective, which increases the levels of excitation of both the optical phonons and carriers, and thus their non-equilibrium temperature increases while the carriers' drift velocity decreases.

Figure 24 shows the average energy of the carriers in steady state as a function of electric field intensity. The results of the Monte Carlo simulation represent the study by Kolník *et al* [135]. In the NESEF calculation, the average energy is given by $3k_\mathrm{B}T_e^*/2 + m_e^* v_e^2/2$, where $T_e^*$ is the electron non-equilibrium temperature. The agreement is good except under high fields, where the difference is less than 8%, which can be ascribed to the fact that intervalley scattering (not considered in the NESEF calculations) may begin to have some influence.

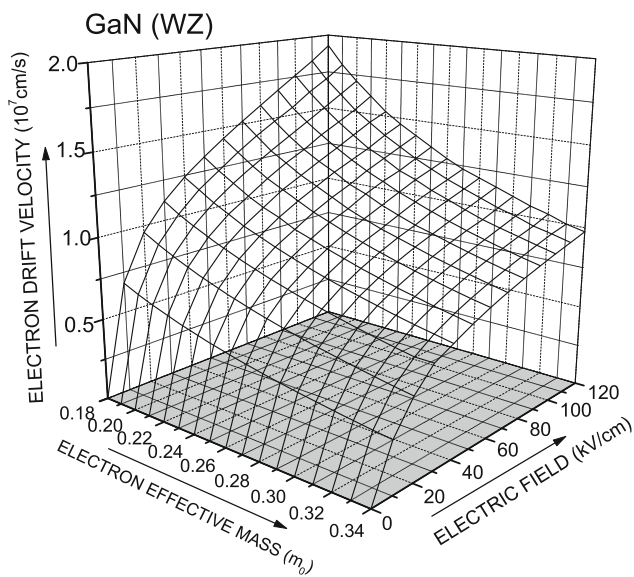

**Figure 22.** Electron drift velocity vs. electron effective mass in steady state of n-GaN (WZ), for several values of electric field.

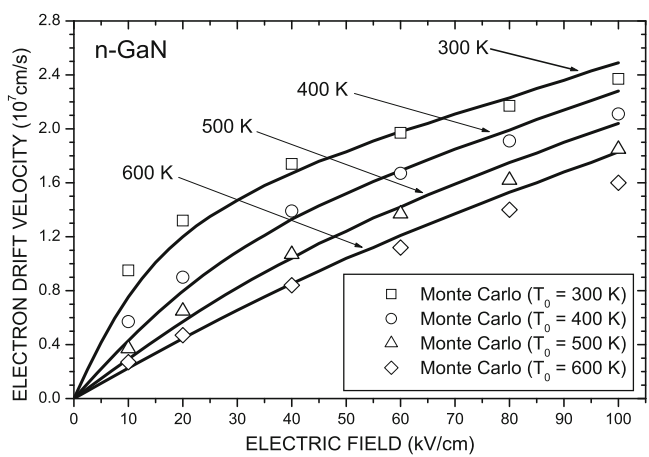

**Figure 23.** Electron drift velocity vs. electric field intensity in n-GaN, comparing the result of NESEF-based kinetic theory (full line) with Monte Carlo simulation [134], when $n = 10^{17}$ cm$^{-3}$ and $T_0 = 300$ K, $T_0 = 400$ K, $T_0 = 500$ K, $T_0 = 600$ K.

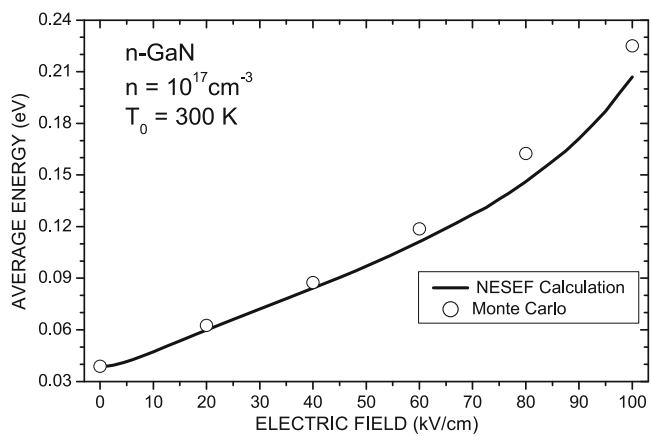

**Figure 24.** Carriers' mean energy vs. electric field intensity in steady state of n-GaN, comparing the result of NESEF-based calculations (full line) with the Monte Carlo simulation [135].

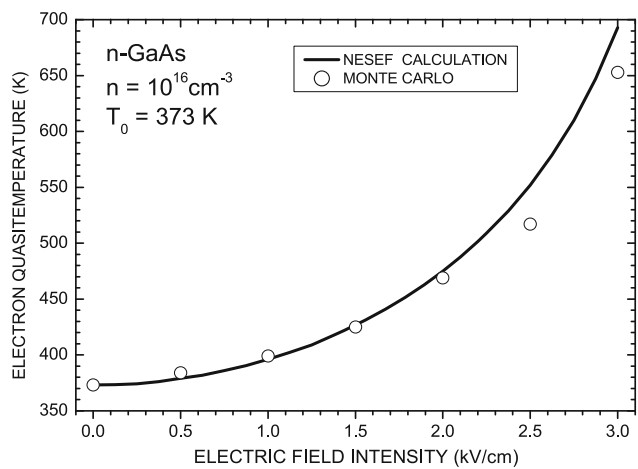

**Figure 25.** Electron quasitemperature vs. electric field intensity, in steady state of n-GaAs, comparing the result of NESEF-based kinetic theory (full line) with the Monte Carlo simulation [136].

Now consider the case of doped GaAs, a material for which experimental results (aside from Monte Carlo simulations) are available.

Figure 25 demonstrates the dependence of the electron non-equilibrium temperature $T_e^*$ on the electric field in the steady state. NESEF results are compared with those of Hilsum [136], who determined the distribution of the carriers using Monte Carlo simulations and then derived the non-equilibrium temperature through fitting to a Maxwell–Boltzmann-like behaviour. The agreement between the two is quite good.

Ruch and Kino [137] reported measurements of the drift velocity of carriers in the steady state of n-GaAs, when $n = 10^{18}$ cm$^{-3}$ and $T_0 = 300$ K. These results are compared in figure 26 with those calculated using NESEF, indicating that there is excellent agreement for values of electric field up to $\sim 2.5$ kV/cm, corresponding to an Ohmic regime. For fields of the order of $\sim 3$ kV/cm and above, intervalley scattering, which is not considered in the NESEF calculations, is responsible for the nonlinear behaviour evidenced in the experiment.

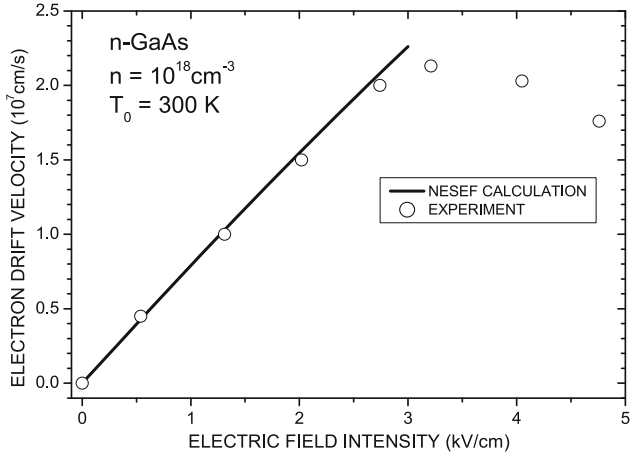

**Figure 26.** Electron drift velocity vs. electric field intensity, in steady state of n-GaN, comparing the result of NESEF-based kinetic theory (full line) with experimental data [137]. There is a good agreement at low fields, and no comparison is possible for electric fields larger than 2.5 kV/cm because the calculations do not include intervalley scattering.

Figure 27a considers the case of n-GaAs with $n = 5 \times 10^{15}$ cm$^{-3}$, where the mobility is evaluated as a function of the reservoir temperature for an electric field of 1 kV/cm. The full line shows the calculations using NESEF, which compares well with the experimental results obtained by Bolger *et al* [138], Hicks and Manley [139] and Rode and Knight [140]. Very good agreement is observed up to 500 K. After that, differences of less than 20% at 900 K are observed. This can be ascribed to the fact that a unique relaxation time for anharmonic processes has been used in the NESEF calculation, which is the one expected at 300 K, but is larger than expected at 900 K. At higher temperatures, relaxation processes are impaired, and the excited optical phonons will cause shorter relaxation times of the momentum, leading to lower conductivity.

The same analysis in the previous paragraph can be applied now in the case of p-doped GaAs, with $p = 10^{16}$ cm$^{-3}$ and an electric field of 1 kV/cm. The experimental data shown in figure 27b were published by Wiley and DiDomenico [141]. The NESEF calculation compares well with the observation, in the interval between 250 K and 350 K; at 400 K, the difference is of the order of 10%, and is slightly larger at 200 K. Again, as in the previous case, this can be attributed to using a unique relaxation time to account for anharmonic effects (at 300 K) over the entire temperatures range.

It can be observed in figure 28 that there exists good qualitative and semiquantitative agreement between the theoretical and experimental results based on the temperature dependence of the hole mobility in cubic GaN [56] under the conditions of the experiment conducted by [142]. The differences observed at the lower and the higher temperatures can be ascribed to indeterminate factors in the theoretical calculation due to: (i) inaccuracies related to the value of hole effective mass [143,144], (ii) at low temperatures, inaccuracy in the value of density of impurities (the corresponding scattering operator is sensitive to this density); however, it can be shown that if $6 \times 10^{18}$ cm$^{-3}$ is used instead of the reported estimated value of $4 \times 10^{18}$ cm$^{-3}$, the agreement becomes better; (iii) at high temperatures, scattering due to the Fröhlich interaction dominates, which in the calculations has been used in the form of a bare Fröhlich potential. However, at the hole concentrations involved, screening effects may become relevant, and then the resulting mobility would increase, leading to a better agreement with the experimental results; (iv) finally, we assumed that Hall scattering factor $r_H = 1$ in the expression $\mathcal{M}_H = r_H \mathcal{M}_c$, where $\mathcal{M}_H$ is the experimental Hall data mobility, when comparing the theoretical results with the experimental data. If the weak temperature dependence of the Hall factor is considered (a variation of the order of 10%), better agreement can be obtained, because $r_H$ can be considerably higher than 1 at low temperatures, but is very close to 1 at high temperatures [145].

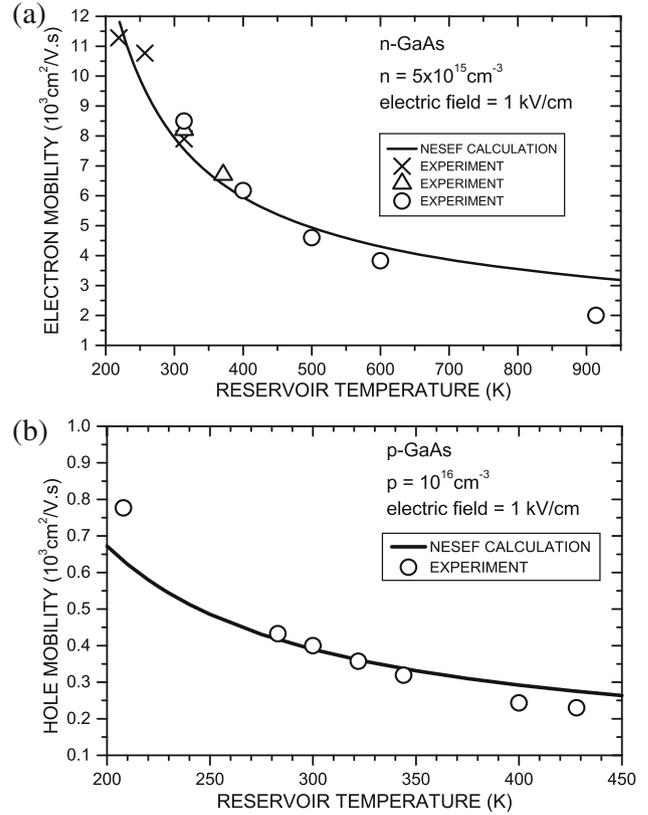

**Figure 27.** (**a**) Electron mobility in steady state of n-GaN for different values of reservoir temperature, comparing NESEF-based calculations (full line) with experimental data (× from ref. [139], ○ from ref. [140], △ from ref. [138]) and (**b**) hole mobility in steady state of n-GaN for different values of reservoir temperature, comparing NESEF-based calculations with experimental data (from ref. [141]).

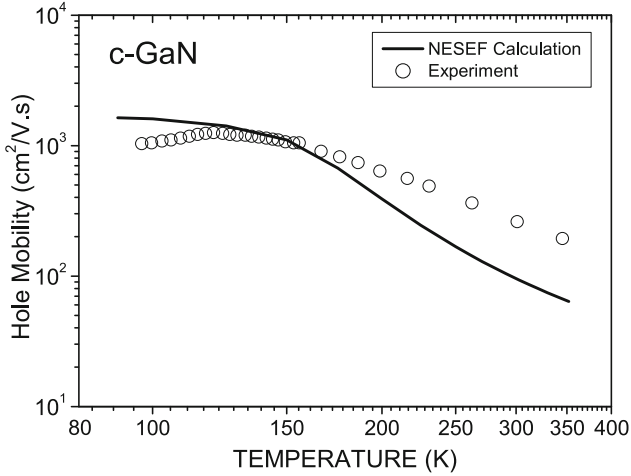

**Figure 28.** Temperature dependence of mobility in ZB GaN in the conditions of the experiment in ref. [142]. Full line is NESEF-based theoretical mobility and dots are experimental data from ref. [142].

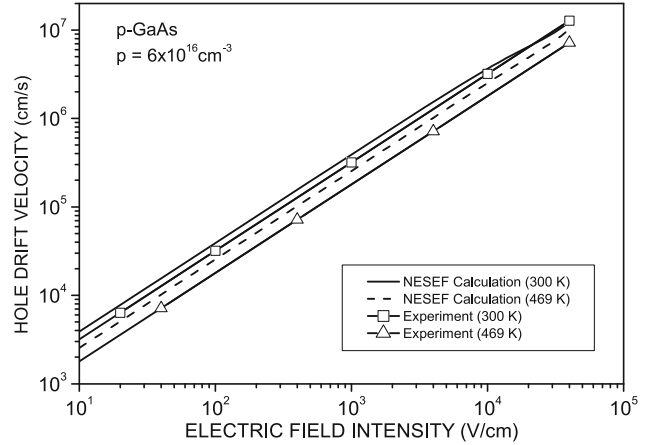

**Figure 29.** Drift velocity of holes in p-GaAs for a broad set of values of electric field intensity, and two values of reservoir temperature, comparing NESEF-based calculations with experimental data ($\square$ and $\triangle$ from ref. [146]).

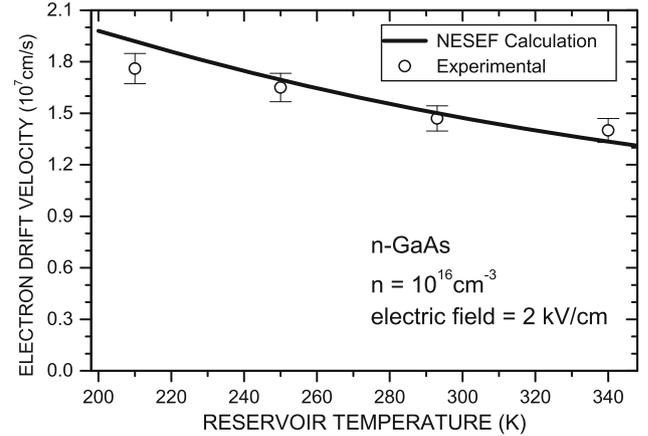

**Figure 30.** Drift velocity of electrons in steady state of n-GaN, for different values of reservoir temperature, comparing NESEF-based calculations (full line) with experimental data from ref. [137].

Figure 29 provides a comparison with the experimental data reported by Dalal and Dreeben [146] for p-doped GaAs, with $p = 6 \times 10^{18}$ cm$^{-3}$, which analyses the drift velocity of the holes in the steady state over a wide range of values of the electric field (no side valleys are present in the valence band), and for two reservoir temperatures, 300 K and 469 K. The agreement is good within the experimental error for $T_0 = 300$ K, and differences of the order of 10% are observed when $T_0 = 469$ K, which are tentatively ascribed to the causes pointed out in the previous paragraph.

Figure 30 shows the drift velocity of electrons in steady state for n-doped GaAs, with $n = 10^{16}$ cm$^{-3}$ and an electric field of 2 kV/cm, as temperature of the reservoir changes. The experimental points are obtained from Ruch and Kino [137] and the full line indicates the NESEF-based calculation. Overall agreement is good except at the lowest temperatures, where the difference is approximately 8%, which falls outside the experimental error.

In the case of n-doped GaN, figure 31 presents a multiple comparison, namely, the drift velocity obtained from experimental data measured by Wraback *et al* [25], a Monte Carlo computation performed by Kolník *et al* [135] and our NESEF results. Some disagreement can be observed among the three results: Monte Carlo values are $\sim 60\%$ higher than the experimental values, while they are $\sim 40\%$ higher for the NESEF results. However, the increasing trend observed among all results compares well. Note that the velocity in the experiment was determined via the change in the electroabsorption associated with the carrier transport using a fitting based on Franz–Keldysh effect (eq. (2) in ref. [25]), which is somewhat imprecise. An alternative, and apparently more precise, experiment using optical measurements has been suggested – Raman scattering of carriers under an electric field – that can also allow for measurements in the transient regime (femtosecond scale) [147].

We next consider studies of the transient regime.

Consider n-GaAs, with $n = 10^{15}$ cm$^{-3}$, $T_0 = 300$ K, and an electric field of 1 kV/cm. Figure 32 shows the evolution of the electron drift velocity, which does not exhibit a velocity overshoot. The agreement between both calculations is good, although the Monte Carlo simulation by Ruch [148] yields larger values (with differences less than 6%) than those obtained by NESEF.

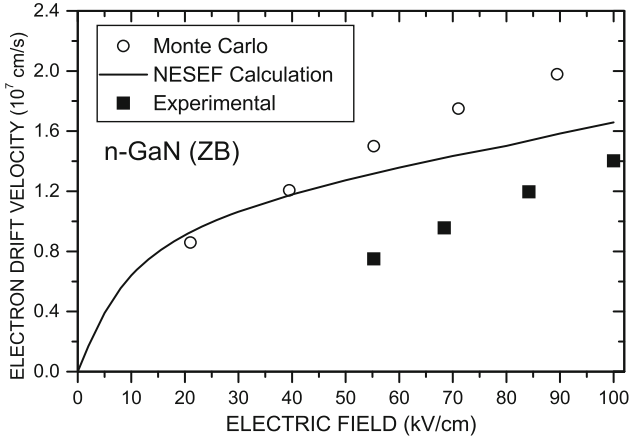

**Figure 31.** Steady-state electron drift velocity as a function of electric field in n-doped ZB GaN when $T_0 = 300$ K, $n \simeq 5 \times 10^{18}$ cm$^{-3}$. Circles: Monte Carlo calculation from [135], squares: experimental data from [25].

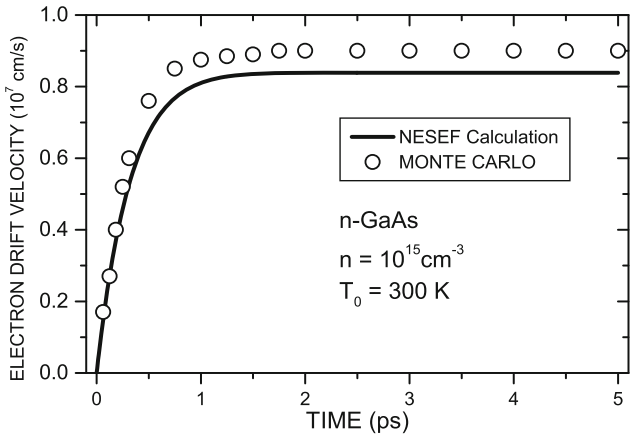

**Figure 32.** Evolution of drift velocity of electrons in n-GaN, comparing NESEF-based calculations (full line) with Monte Carlo simulation of ref. [148].

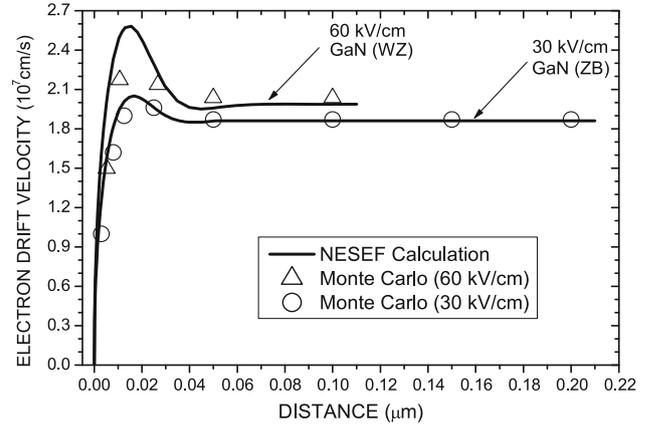

**Figure 33.** Evolution of drift velocity of electrons in n-GaN in terms of travelled distance, for an electric field intensity of 30 kV/cm (for ZB GaN) and 60 kV/cm (for WZ GaN), comparing NESEF based calculations (full curve) with Monte Carlo simulation [38].

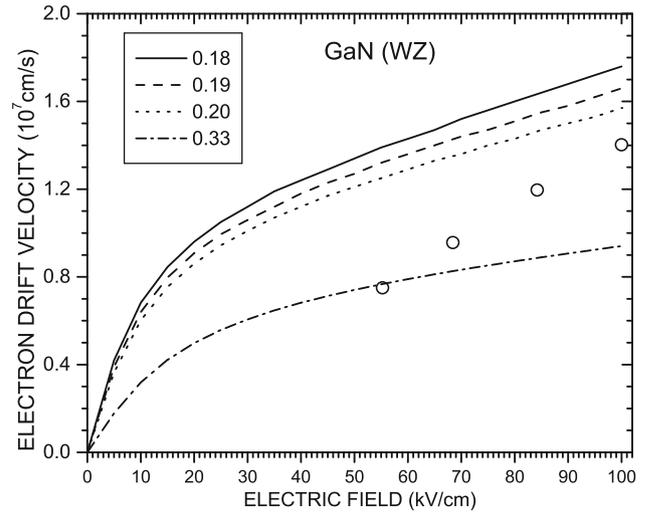

**Figure 34.** Steady-state electron drift velocity as a function of electric field: $m_e^* = 0.18m_0$, ref. [122]; $m_e^* = 0.19m_0$, ref. [103]; $m_e^* = 0.20m_0$, ref. [123]; $m_e^* = 0.33m_0$, ref. [124], where $m_0$ is the electron rest mass. Circles: experimental values from [5].

Next, consider zincblende n-GaN, with $n = 10^{17}$ cm$^{-3}$ and $T_0 = 300$ K, and the time evolution of the drift velocity of the carriers in the presence of an electric field of intensity 30 kV/cm. Figure 33 compares the NESEF results with those of a Monte Carlo simulation conducted by Foutz *et al* [38]. The horizontal axis is expressed in terms of the carrier's travel length, and the values shown correspond to an elapsed time of approximately 1 ps (travel distance of 0.2 $\mu$m). Both calculations predict a velocity overshoot, and their agreement is very good. A similar situation is also considered in figure 33 at a field intensity of 60 kV/cm, except that the crystalline structure is wurtzite. A velocity overshoot is also predicted, and the agreement between both results is good, but not as good as in the case of GaN (ZB). The differences, which are less than 10%, may be attributed to the fact that the average electron effective mass has been used in the NESEF-based calculations and not the asymmetric mass present in wurtzite structures.

Similar to figure 7, figure 34 shows the steady-state electron drift velocity as a function of electric field for four different values of electron effective mass [103, 122–124] for GaN (WZ). The circles in figure 34 represent the measurement values for GaN obtained by Wraback *et al* [25].

**Table 3.** Comparison of the electron drift velocity ($\times 10^7$ cm/s) in n-doped InN obtained in this work with other results.

| $\mathcal{E}$ (kV/cm) | This work ($m_e^* = 0.045m_0$) | Ref. [129] ($m_e^* = 0.045m_0$) | Ref. [149] ($m_e^* = 0.07m_0$) |
|---|---|---|---|
| 5 | 2.6 | 3.0 | 0.5 |
| 10 | 3.5 | 4.2 | 0.9 |
| 15 | 4.2 | 5.0 | 1.3 |

To conclude this section, table 3 shows the comparison of electron drift velocity in n-doped InN obtained in this study with other results [129,149] for three different electric field intensities. The calculations were performed under conditions similar those in [129,149], that is, the lattice temperature was 300 K and the free electron concentration was $1 \times 10^{17}$ cm$^{-3}$. It can be observed that the obtained results are in good agreement with the values obtained by O'Leary and Foutz [129] and with the lowest electron drift velocity obtained by Masyukov and Dmitriev [149]. This can be explained by the fact that in this study and in [129], the researchers used an electron effective mass of $0.045m_0$, while Masyukov and Dmitriev used an electron effective mass of $0.07m_0$ (as expected, the electron drift velocity decreases with increasing electron effective mass).

## 4. Concluding remarks

The summarised results presented in this paper, focussing on n-doped GaN, InN, AlN and GaAs in moderate to strong electric fields, are derived from the framework provided by NESEF, and provide an extensive view of their transport properties. The main points that can be highlighted are as follows:

(1) In the case under consideration, the NESEF-based kinetic theory has been used in the Markovian approximation (i.e., up to second-order interaction strengths). Below this limit, the collision operators $J^{(2)}$ acquire the form of the Golden Rule of Quantum Mechanics. However, they are averaged over the non-equilibrium NESEF ensemble and depend on time as a result of the time evolution of the non-equilibrium thermodynamic variables. This dependence on time results because irreversible processes are developing in the system while it is probed. In the steady state, there is no change in time because dynamical equilibrium is established (but with entropy production) in the pumping and relaxation effects.

(2) In the transient regime, an overshoot occurs in drift velocity. It occurs only at intermediate to high field intensities, and is more pronounced when such intensity is greater. A less-pronounced overshoot in non-equilibrium temperature was also observed.

(3) The transient time that elapses until a steady state is attained is very short, on a scale of a few hundred femtoseconds. It decreases as the field intensity increases, but only slightly.

(4) In the steady state, a near-Ohmic regime is present at low fields, as expected, but a nonlinear law occurs at intermediate to high fields. In the Ohmic regime, the non-equilibrium temperature approximately increases with the square of the electric field intensity.

(5) In the non-Ohmic regime at intermediate to high fields, while the non-equilibrium temperature maintains a square-law in the field (but with linear corrections arising out of the dependence on the field of the relaxation times), the drift velocity (and the current) exhibits a strong departure from linearity with respect to the field intensity. This departure has the characteristic that the drift velocity (or the current) increases more slowly with increasing field intensity. In other words, the field-dependent differential conductivity defined as $dJ_e/d\mathcal{E}$ (where $J_e = -nev_e(t)$) continues decreasing as the electric field $\mathcal{E}$ is increases.

(6) It should be noted that a parabolic band approximation (effective mass approximation) was used in the NESEF calculations, which yields valid calculations up to a certain value of the electric field (the end points in the lines in the figures for each compound). For higher field intensities, intervalley scattering begins to play a role. This effect can be incorporated into the theory with some additional effort (the equations would be modified to include the effects of all the valleys and intervalley scattering rates).

(7) The non-equilibrium temperature of the LO phonons weakly increases (up to a 10% increase at the highest fields considered). This occurs because the excess energy acquired by the $10^{17}$cm$^{-3}$ carriers is redistributed among all the phonon modes in the Brillouin zone. A more detailed description of the non-equilibrium macroscopic state of the system would consist of introducing the populations of the LO phonons per mode (instead of their global energy), and then the intensive non-equilibrium thermodynamic variable would represent the non-equilibrium temperature per mode. In this case, it would reflect the so-called 'hot-phonon temperature overshoot' [55,150,151].

(8) The non-equilibrium temperature of the acoustic phonons is practically unaltered (at most an increase of 2% at the highest fields considered). This is a consequence of the fact that the anharmonic interaction and the carrier–AC phonon interaction are weak. Then, the carriers produce a slow rate of excess energy transfer to the AC phonons, which is rapidly dissipated by the process of heat diffusion to the reservoir (provided that there is good thermal contact between the sample and the reservoir).

(9) With increasing intensity of the electric field, the mean energy of the carriers increases (the latter consists of kinetic thermal energy, roughly $3k_B T_e^*/2$, plus the kinetic energy of drift, $m_e^* v_e^2/2$). For all compounds that have been considered (and we argue that it is a general rule) in the steady state, the thermal kinetic energy is always greater than the kinetic energy of drift. In other words, the pumped energy is distributed so that thermal energy is preferred (thermal chaotic movement) to kinetic energy because of the drift motion.

(10) Under the conditions imposed in the calculations (the 'thought-of experimental protocol'), scattering caused by impurities is negligible compared to the scattering of carriers by phonons, even in comparison with the weak scattering involving AC phonons. As expected, the mobility (of the order of a few hundred cm$^2$/V · s) is a result of the strong dominance of scattering via the Fröhlich potential of carriers and LO phonons in these strong polar semiconductors.

(11) The highest mobility corresponds to InN, which can be attributed to the fact that its electrons have a lower effective mass (see table 1) than electrons in GaN and AlN, and their mobility $\mathcal{M}$ follows the $\mathcal{M}_{InN} > \mathcal{M}_{GaN} > \mathcal{M}_{AlN}$ hierarchy.

(12) We compared the NESEF results with the results of computational experiments, namely Monte Carlo simulations, and the results indicated very good agreement between the two approaches.

(13) Seemingly, the advantage of NESEF over Monte Carlo that the formalism, as discussed, provides a set of nonlinear equations of evolution for the basic non-equilibrium thermodynamic variables, and the transport properties obtained, provide a clear picture of the physical phenomena that develop in the non-equilibrium (dissipative) system. On the one hand, the kinetic equations make it clear which microscopic (quantum dynamics) processes contribute to the time-dependent ensemble average (statistical formulation) and on the other hand, a set of coupled nonlinear integro-differential equations that allow relatively accessible mathematical (computational) handling are derived using the NESEFapproach.

(14) We stress that it is important to use an accurate value of the electron effective mass when performing theoretical studies and simulations of electronic transport in III nitrides (GaN, AlN and InN), because the electron drift velocity and electron non-equilibrium temperature depend critically on the value assumed for the electron effective mass (see figures 7–10, 21, 22, 34, and tables 2 and 3).

## Appendix A. NESEF kinetic theory in brief

The non-equilibrium statistical ensemble formalism (NESEF) is extensively described in several publications, for example refs [94,152–158]. A condensation of its basic tenets and construction steps, as well as the discussion of shortcomings and criticisms, is presented in [43,159]. The first fundamental steps are:

*Step* 1: The separation of the total Hamiltonian of the system into two parts, namely

$$\hat{H} = \hat{H}_0 + \hat{H}', \tag{A.1}$$

where $\hat{H}_0$ (the so-called 'secular' or 'quasi-conserving' part) consists of the kinetic energies and the interactions strong enough to produce damping of correlations with relaxation times smaller than the characteristic time associated with the experiment (typically the resolution time of the detection apparatus). On the other hand, the 'relaxing' or 'non-conserving' part contains the remaining interactions, leading to relaxation processes with lifetime larger than the afore-mentioned characteristic time; and also the energy operators of interaction of system and external pumping sources.

*Step* 2: A selection rule for the choice of the basic set of dynamical variables, say $\{\hat{P}_j(\mathbf{r})\}$, $j = 1, 2, \ldots$, which must satisfy the condition that [160]

$$\frac{1}{i\hbar}[\hat{P}_j(\mathbf{r}), \hat{H}_0] = \sum_k \Omega_{jk} \hat{P}_k(\mathbf{r}). \tag{A.2}$$

Coefficients $\Omega_{jk}$ can be, depending on the case, c-numbers, differential operators, or, eventually space-dependent functions. These two steps work in unison, in the sense that in this way, fast and slow relaxing variables are separated.

*Step* 3: Construction of the statistical operator in terms of the set above. Let us call it $\hat{\rho}(t)$, and introduce the set of accompanying macrovariables

$$Q_j(\mathbf{r}, t) = \text{Tr}\{\hat{P}_j(\mathbf{r})\hat{\rho}(t)\}, \tag{A.3}$$

which evidently are the averages of $\hat{P}_j(\mathbf{r})$ over the non-equilibrium ensemble. Next, the statistical operator is constructed either using heuristic arguments [152], or, better, using the variational method [153–156]. The statistical operator is given in Zubarev's approach [152–154] by

$$\hat{\rho}_\varepsilon(t) = \exp\left\{ \ln \hat{\bar{\rho}}(t, 0) \right.$$
$$\left. - \int_{-\infty}^{t} dt' e^{\varepsilon(t'-t)} \frac{d}{dt'} \ln \hat{\bar{\rho}}(t', t'-t) \right\}, \tag{A.4}$$

where

$$\hat{\bar{\rho}}(t, 0) = \exp\left\{ -\phi(t) - \sum_j d^3r\, F_j(\mathbf{r}, t)\hat{P}_j(\mathbf{r}) \right\}, \tag{A.5}$$

$$\hat{\bar{\rho}}(t', t'-t) = \exp\left\{ -\frac{(t'-t)\hat{H}}{i\hbar} \right\} \hat{\bar{\rho}}(t', 0)$$
$$\times \exp\left\{ \frac{(t'-t)\hat{H}}{i\hbar} \right\} \tag{A.6}$$

and the set $\{F_j(\mathbf{r}, t)\}$ is constituted by the Lagrange multipliers that the method introduces. An important point to stress is that these Lagrange multipliers define a complete set of intensive non-equilibrium thermodynamic variables, which fully describe the macroscopic state of the system, as, alternatively, do the macrovariables $Q_j(\mathbf{r}, t)$. Equation (A.3) relates both sets of variables. Moreover, $\phi(t)$ ensures the normalisation of both the distribution $\hat{\rho}_\varepsilon(t)$ and the auxiliary distribution $\hat{\bar{\rho}}(t, 0)$ and $\varepsilon$ is a positive infinitesimal that goes to zero after the trace operator in the calculations of averages has been performed.

*Step* 4: Construction of a non-equilibrium statistical ensemble formalism-based kinetic theory, which follows straight-forwardly, in that the equations of evolution for the basic variables (describing the irreversible evolution of the macrostate of the system) are the average over the non-equilibrium ensemble of Heisenberg equations of motion. That is,

$$\frac{\partial}{\partial t} Q_j(\mathbf{r}, t) = \text{Tr}\left\{ \frac{1}{i\hbar}[\hat{P}_j(\mathbf{r}), \hat{H}]\,\hat{\rho}_\varepsilon(t) \right\}, \tag{A.7}$$

which can be rewritten as

$$\frac{\partial}{\partial t} Q_j(\mathbf{r}, t) = J_j^{(0)}(\mathbf{r}, t) + J_j^{(1)}(\mathbf{r}, t) + \mathfrak{F}_j(\mathbf{r}, t), \tag{A.8}$$

where

$$J_j^{(0)}(\mathbf{r}, t) = \text{Tr}\left\{ \frac{1}{i\hbar}[\hat{P}_j(\mathbf{r}), \hat{H}_0]\hat{\rho}(t) \right\}, \tag{A.9}$$

$$J_j^{(1)}(\mathbf{r}, t) = \text{Tr}\left\{ \frac{1}{i\hbar}[\hat{P}_j(\mathbf{r}), \hat{H}']\hat{\rho}(t) \right\}, \tag{A.10}$$

$$\mathfrak{F}_j(\mathbf{r}, t) = \text{Tr}\left\{ \frac{1}{i\hbar}[\hat{P}_j(\mathbf{r}), \hat{H}']\hat{\rho}'_\varepsilon(t) \right\}, \tag{A.11}$$

following from the fact that using eq. (4) it can be written as

$$\hat{\rho}_\varepsilon(t, 0) = \hat{\bar{\rho}}(t, 0) + \hat{\rho}'_\varepsilon(t, 0), \tag{A.12}$$

with details given in [80,161].

Finally, NESEF provides microscopic (mechano-statistical) foundation to irreversible thermodynamics, in the form of Informational Statistical Thermodynamics [47].